\documentstyle[12pt]{article}

\columnwidth\textwidth

\begin{document}

\newcommand{\be}{\begin{equation}}
\newcommand{\ee}{\end{equation}}
\newcommand{\beann}{\begin{eqnarray*}}
\newcommand{\eeann}{\end{eqnarray*}}
\newcommand{\bea}{\begin{eqnarray}}
\newcommand{\eea}{\end{eqnarray}}
\newcommand{\nn}{\nonumber}
\newcommand{\ben}{\begin{enumerate}}
\newcommand{\een}{\end{enumerate}}
\newtheorem{df}{Definition}
\newtheorem{thm}{Theorem}
\newtheorem{lem}{Lemma}
\newtheorem{prop}{Proposition}
\begin{titlepage}

\noindent
\hspace*{11cm} BUTP-97/7 \\
\vspace*{1cm}
\begin{center}
{\LARGE Group-Theoretical Quantization of 2+1 Gravity in the
Metric-Torus Sector}    

\vspace{1cm}

P. H\'{a}j\'{\i}\v{c}ek \\
Institute for Theoretical Physics \\
University of Bern \\
Sidlerstrasse 5, CH-3012 Bern, Switzerland \\
\vspace*{1cm}

May 1997 \\ \vspace*{.5cm}

\nopagebreak[4]

\begin{abstract}
A symmetry based quantization method of reparametrization invariant
systems is described; it will work for all systems that possess
complete sets of perennials whose Lie algebras close and which
generate a sufficiently large symmetry groups. The construction leads
to a quantum theory including a Hilbert space, a complete system of
operator observables and a unitary time evolution. The method is
applied to the 2+1 gravity. The paper is restricted to the
metric-torus sector, zero cosmological constant $\Lambda$ and it makes
strong use of the so-called homogeneous gauge; the chosen algebra of
perennials is that due to Martin. Two frequent problems are
tackled. First, the Lie algebra of perennials does not generate a
group of symmetries. The notion of group completion of a
reparametrization invariant system is introduced so that the group
does act; the group completion of the physical phase space of our model
is shown to add only some limit points to it so that the ranges of
observables are not unduly changed. Second, a relatively large number
of relations between observables exists; they are transferred to the
quantum theory by the well-known methods due to Kostant and
Kirillov. In this way, a uniqueness of the physical representation of
some extension of Martin's algebra is shown. The Hamiltonian is
defined by a systematic procedure due to Dirac; for the torus sector,
the result coincides with that by Moncrief. The construction may be
extensible to higher genera and non-zero $\Lambda$ of the 2+1 gravity,
because some complete sets of perennials are well-known and there are
no obstructions to the closure of the algebra.

\end{abstract}

\end{center}

\end{titlepage}

\section{Introduction}
\label{sec:intro}
The (2+1)-dimensional gravity is a popular model. It has been utilized
as a laboratory for studies of specific problems in quantum gravity
(for reviews, see \cite{carlip1}, \cite{loll1}). Our aim is to
illustrate a symmetry based method of quantization of the so-called
\textit{reparametrization invariant systems} (RIS); a RIS is a
constrained system whose dynamics is completely determined by its
constraints (the value of the Hamiltonian is zero). For this purpose,
2+1 gravity seems particularly interesting, because it is a generally
covariant RIS and the structure of its constraints is very similar
to that of the general relativity.

The (technical) starting point of the method is a choice of a set of
functions on the phase space that satisfy three conditions. First,
they should have vanishing Poisson brackets with all constraints. Such
quantities have been called ``first-class'' by Dirac \cite{dirac1} and
they are often called 
``observables'' today. The name ``observables'' is, however, not
justified in the case of RIS's. This difficulty, which is connected to
the problem of time, has been first noticed by Kucha\v{r}
\cite{cordoba}, who has introduced a special name,
\textit{perennials}, for the first-class quantities of RIS's. An
attempt at a mathematical formulation of Kucha\v{r}'s ideas 
can be found in \cite{PH5}, where a construction of observables from
perennials is described that necessarily makes use of some additional
(time) structures (cf.\ Sec.\ \ref{sec:particle})). The second
condition on the set is that it separates the classical
solutions. This means that for two different solution there is at
least one function from the set that takes on different values at
these solutions. This property is called \textit{completeness}, it was
first introduced by Bergmann \cite{bergmann1} and it was later used by
Ashtekar \cite{bluebook} as one of the basic properties within the
algebraic quantization method. The third condition on the set is that
it forms a closed algebra with respect to linear combinations and
Poisson brackets. Thus, it will be a Lie algebra $\mathbf{g}$.

However, availability of perennials in the general relativity, or even
their existence, has been questioned \cite{cordoba}. Indeed, one \textit{can}
invent RIS's that possess no perennials. However, these are
mathematical constructs whose physics is strange: they cannot be
reduced, even locally, and their physical phase spaces are not
manifolds. It turn out that physically reasonable systems do possess
complete systems of perennials (for proof, see \cite{PH5}). True,
perennials of a special kind, for example those that are linear in
momenta or are local in certain sense are already shown to be absent
in the general relativity \cite{kuchS}, \cite{AT}. We shall show,
however, how these proofs are transcended for the 2+1 gravity model
(to which the proof methods of Kucha\v{r}, Anderson and Torre may be
applicable). The third condition, that the perennials form a Lie
algebra, might also be a source of problems; there are finite
dimensional symplectic manifolds that do not admit a complete finite
set of perennials whose algebra closes. However, it seems that such
systems are again rather artificial with
strange physics and that they also would be very difficult to quantize
by any existing method.

The complete algebra of perennials plays a twofold r\^{o}le in the
theory. First, as already mentioned, observables can be constructed
from them. We assume that these observables comprise the most
important and directly measurable properties of the system---this
condition, although a little vague one from the mathematical point of
view, should influence the choice of the algebra
$\mathbf{g}$.  Second, a group $G$ of symmetries is generated by
$\mathbf{g}$, if some conditions are satisfied. Such a group, if it
exists, can strongly simplify the task of finding suitable
representation of the algebra $\mathbf{g}$ (that is, quantizing the
system). The corresponding methods are those of the so-called
\textit{group quantization} (see \cite{ishamG} for a review); a
modification of the methods suitable for the RIS's has been suggested
by Rovelli \cite{R-group}. The existence of the group $G$ can however
also help to solve the problem of time that afflicts the quantization
of RIS's. The idea of this use for $G$ stems from an old paper by
Dirac \cite{dirac2}; in this paper, a time evolution has been
constructed for a system of relativistic particles on the Minkowski
spacetime. Dirac's ideas have been extended to general RIS's and
developed to a coherent theory in \cite{PH1} and \cite{PH5}.  Although
already published, this theory is far from being well-known. We give,
therefore, a short pedagogic exposition in Sec.\ \ref{sec:particle} to
make the paper self-contained.

The group $G$ of symmetries is obtained from the algebra $\mathbf{g}$
in two steps. First, each Lie algebra determines a unique (simply
connected) abstract Lie group $G$. Second, the Hamiltonian vector
fields of the elements of the algebra determine an action of $G$ on
the phase space, if the vector fields are complete---we shall meet a
prominent example of incomplete Hamiltonian vector fileds in the 2+1
gravity model. The action is then unique; let us call it
\textit{Hamiltonian action} of $G$. In some cases, the center of $G$
will contain a non-trivial subgroup $G_c$, whose elements act
trivially. Two cases must be distinguished: isolated elements of $G$
in $G_c$ and Lie subgroups of $G$ in $G_c$. One can simplify $G$ by
taking the quotient with respect to the isolated elements, but not
with respect to the Lie part of $G_c$ \cite{ishamG}.

The 2+1 gravity model can be considered as a dynamics of an ISO(1,2)
affine connection of a three-dimensional manifold $M$
\cite{carlip}. If some conditions are satisfied, then the affine
connection defines a Lorentzian metric on $M$ and this property is
preserved by the dynamics (we shall cut out the singularities!). We
assume that Cauchy surface $\Sigma$ is the torus ($S^1\times S^1$) and
that the metric is well-defined; in such a case, we speak of the 
\textit{metric-torus sector}. This sector alone gives a solvable but
unexpectedly interesting model so that all the general points above
can be illustrated in a rather non-trivial way. We shall limit
ourselves to this sector and we will work in a particular gauge, the
so-called homogeneous gauge, in which the three-metric depends only on
the time coordinate. Let us remark that there is no reason why our
method should not work for higher genera. One can try to use the loop
variables described in \cite{mart} or \cite{loll2}; the topology of
the physical phase spaces for all higher genera is contractible and
there are global Darboux coordinates; thus, there are no obstructions
to the existence of a complete system of functions with a closed
algebra.

In the present paper, we will choose the complete set of perennials
that has been published by Martin \cite{mart}. The perennials are
directly related to the topological degrees of freedom of the system
and can be expressed as a kind of Wilson loop variables. For the
metric-torus sector, Moncrief observed that Martin's perennials form
the six-dimensional algebra ${\mathbf g}$ isomorph to iso(1,2). We
shall introduce new canonical coordinates that are adapted to Martin's
perennials; the constraint becomes formally the mass-shell condition
for a rest-mass-zero relativistic particle. Thus, further four
perennials can be immediately written down; together with the old
ones, they form a ten-dimensional algebra so(2,3). The corresponding
group $G$ is the conformal group of three-dimensional flat spacetime;
$G$ is isomorph to SO(2,3).

The problem of existence of the Hamiltonian action for $G$ is
non-trivial. Our calculation will reveal that only a subgroup $G_0
\subset G$ with the structure of SO(1,2)$\times{\mathbf R}$ acts on the
phase space of the system. It is easy to observe, however, that the
phase space can be extended so that the whole group $G$ has the
Hamiltonian action on the extended space. We call such extensions 
\textit{group completions}. A minimal group completion is unique under some
quite general conditions. It turns out that the minimal group
completion of the physical phase space consists, in our case, of adding
``relatively few points'' in such a way that the ranges of observables
are not changed except for adding some boundary points to them.

The metric-torus sector has two degrees of freedom. Thus, a
ten-dimensional algebra like ${\mathbf g}$ will exhibit six
independent relations. Using the Kostant-Kirillov method, we find that
the group $G$ does not possess any physical representation.  This has
to do with Van Hove theorem: the physical phase space is too small and
the algebra $\mathbf g$ of functions is to large to be represented
without deformation. We find, however, that one of its maximal
subgroups, $G_1$, which is seven-dimensional, and isomorph to
(SO(1,2)$\times{\mathbf R})\otimes_S{\mathbf R}^3$, possesses a unique
physical representation and we calculate the form of the operators
representing ${\mathbf g}_1$. Here, ``$\otimes_S$'' denotes the
semidirect product of groups. Three independent classical relations
for $G_1$ can be written down in terms of (generalized) Casimir
operators.

For the construction of time evolution \`{a} la Dirac, we can use only
the four-dimensional subgroup $G_0$; then the Hamiltonian action of
$G_0$ on the constraint surface (and so on the classical spacetimes)
provides the interpretation of the corresponding unitary
transformations in the Hilbert space. It turns out that this action
``goes in time direction'' and so a time evolution can be
constructed. The candidates for the Hamiltonian that generates
everywhere time evolution towards future form a three-dimensional
family. Most of these operators are unbounded from below. Thus, the
condition that an operator generates time evolution towards the future
does not necessarily guarantee that the operator has a non-negative
spectrum. However, there is exactly one Hamiltonian that \textit{is}
bounded from below (and it is even non-negative). The positive
Hamiltonian coincides with the Hamiltonian written down by Moncrief
\cite{M2} and the time coincides with the constant mean external
curvature.

Some surprise is that the quantum mechanics we have constructed is not
equivalent to the ``ordinary'' quantum mechanics of the rest-mass-zero
free particle in three dimensional Minkowski spacetime, in spite of
the fact that the algebra of observables is so(2,3) like for the
particle, and that we managed to introduce new variables in which the
constraint coincides with the ordinary mass-shell condition for such a
particle. The explanation is that the global structure of the torus
configuration space is very different from that of the particle (which
is the three-dimensional Minkowski spacetime): the former is only a
subset of the latter, namely the inside of the light cone of the
origin. The points outside of the light cone correspond to timelike
two-surfaces evolving in a spacelike direction (the signature of the
spacetime remains +1).

The plan of the paper is as follows. In Sec.\ 2, we briefly describe
the mathematical apparatus that has turned out advantageous for the
study of RIS's, their observable properties and their time evolution.
No detail and proofs are given, because these can be found in already
published papers \cite{PH1} and \cite{PH5}. In Sec.\ 3, we list our
starting assumptions and equations concerning the 2+1 gravity
model. They are mostly taken over from \cite{M1} and \cite{M2}, where
more detail can be found. One non-trivial but plausible assumption is
that the so-called homogeneous gauge can be chosen in which the model
becomes finite-dimensional (\cite{M1}, \cite{M-M}).

In Sec.\ 4, we study the problem of action of the group, define the
group completion of a RIS and the \textit{weak action} of the
group. We derive the group completion of the model and prove that the
group completed physical phase space contains the original one as an
open dense subset. We define the action of $G$ on the completed
physical phase space, give the form of all observables obtained from
the algebra ${\mathbf g}$ on the physical phase space and list all
independent relations.

Sec.\ 5 describes an application of Dirac's time evolution idea to our
model. We find that the dynamics is much more unique than in the case
of free relativistic particle studied by Dirac (who found three
inequivalent ``forms of relativistic dynamics''): we find only one
``form''. Finally, in Sec.\ 6, after a brief description of the
Kostant-Kirillov method, we derive the physical representation.

\section{Example: the relativistic particle}
\label{sec:particle}
We consider a free relativistic particle of mass $m>0$ on the
four-dimensional Minkowski spacetime with coordinates $x^\mu$ and the
metric $\eta_{\mu\nu} = \mbox{diag}(-1,1,1,1)$. This is one of the
systems studied originally by Dirac \cite{dirac2}. Let the conjugate
momenta be $p_\mu$ and the constraint be ${\mathcal{H}} = (1/2)(p\cdot p
+ m^2)$. We shall use the abbreviation $A\cdot B := \eta_{\mu\nu}A^\mu
B^\nu$ throughout the paper.

The manifold
${\mathbf{R}}^8$ with the coordinates $x^\mu$ and $p_\mu$, and with the
symplectic form $\tilde{\Omega} = dp_\mu \wedge dx^\mu$ will be called
\textit{extended phase space} and denoted by $\tilde{\Gamma}$.
The submanifold $\Gamma$ of $\tilde{\Gamma}$ defined by
${\mathcal{H}} = 0$ and $p_0 < 0$ is the \textit{constraint
surface}. The function $\mathcal{H}$ is the
so-called \textit{super-Hamiltonian}. It defines $\Gamma$, generates
reparametrizations along the particle trajectories and generates the
dynamics:
\be
  \dot{x}^\mu = {\mathcal{N}} \{x^\mu,{\mathcal{H}}\}, \quad
  \dot{p}_\mu = {\mathcal{N}} \{p_\mu,{\mathcal{H}}\},
\label{xdot-pdot}
\ee where $\mathcal{N}$ is an arbitrary function, the so-called
\textit{lapse}.  The arcs that are determined by maximal solutions of
the equations (\ref{xdot-pdot}) will be called \textit{c-orbits}. A
c-orbit will be typically denoted by $\gamma$. c-orbits represent
classical solutions.  The quotient space $\bar{\Gamma} =
\Gamma/\gamma$ will be called \textit{physical phase space}; we will
assume that $\bar{\Gamma}$ is a manifold. In this case, there is a
symplectic form $\bar{\Omega}$ on $\bar{\Gamma}$ that is uniquely
determined by $\tilde{\Omega}$. In our example, $\Gamma$ is
seven-dimensional, c-orbits are one-dimensional, so $\bar{\Gamma}$ is
six-dimensional. The dimension of the physical space is the double of
the physical degrees of freedom.

An important notion is that of \textit{transversal surface}. This is
any submanifold $\Gamma_i$ of $\Gamma$ such that each c-orbit
intersects $\Gamma_i$ exactly once and in a transversal direction (any
vector at an intersection point that is simultaneously tangential to
the c-orbit and to the transversal surface is necessarily the zero
vector). The importance of transversal surfaces for the description of
time evolution in relativistic theories has been recognized by Dirac
\cite{dirac2}. Each transversal surface $\Gamma_i$ carries a unique
symplectic form $\Omega_i$, the pull-back of $\tilde{\Omega}$ to
$\Gamma_i$. The symplectic space $(\Gamma_i,\Omega_i)$ can be
identified with $(\bar{\Gamma},\bar{\Omega})$ by the map that sends
each c-orbit $\gamma$ from $\bar{\Gamma}$ to the intersection point of
$\gamma$ with $\Gamma_i$. An example of transversal surface, which we
denote by $\Gamma_0$, is given by the equations $x^0 = 0$ and
${\mathcal{H}} = 0$. As coordinates on $\Gamma_0$, the functions
$x^1$, $x^2$, $x^3$, $p_1$, $p_2$ and $p_3$ can be chosen; in these
coordinates, $\Omega_0 = dp_k \wedge dx^k$. As we can see, $\Gamma_0$
defines a particular \textit{time instant}, namely $x^0 = 0$. This is
a general property of transversal surfaces; for example, in the
general relativity, a transversal surface defines a unique Cauchy
surface in any generic spacetime solution (ie.\ the solution that
admits no symmetry).

The Poincar\'{e} group (in fact, only the componet of identity
thereof) ISO(1,3) acts on $\tilde{\Gamma}$ in the usual way, leaving
both $\tilde{\Omega}$ and $\Gamma$ invariant. Such transformations in
$\tilde{\Gamma}$ are called \textit{symmetries}. The action is
generated by ten functions $p_\mu$, $J_k := \epsilon_{klm}x_l p_m$ and
$K_k := x_k p_0 - x_0 p_k$, $k = 1,2,3$ (the indices are lowered and
raised by the metric). The functions have vanishing Poisson brackets
with $\mathcal{H}$ (as $\mathcal{H}$ is an invariant), so they are
perennials. They form a Lie algebra that is isomorph to iso(1,3) with
respect to linear combinations and Poisson brackets, and the set
separates c-orbits. Thus, we have an example of a complete Lie algebra
$\mathbf{g}$ of perennials.

In this case, the group that is determined by $\mathbf{g}$ is
Sl(2,$\mathbf{C}$)$ \otimes_S {\mathbf{R}}^4$. The Hamiltonian action
of the group is just the above action of ISO(1,3): The center of
Sl(2,$\mathbf{C}$) contains only one element that is different from
identity, which acts trivially, so we can restrict ourselves to the
group ISO(1,3).

Important general properties of symmetries are (see \cite{R-group} and
\cite{PH1}): a) a symmetry sends c-orbits onto c-orbits and b) it
sends transversal surfaces onto transversal surfaces. The property b)
is crucial to the construction of dynamics \`{a} la Dirac
\cite{dirac2}: by the transformations of the group $G$, a time instant
is sent into another time instant. Dirac also proposed to choose
maximally symmetric (with respect to $G$) transversal surfaces. Such a
choice not only minimizes the number of ``different time instants'',
but also simplifies the operator observables \cite{dirac2}. There are
three inequivalent maximally symmetric transversal surfaces for the
relativistic particle: a) the spacelike plane $x^0 = 0$, b) the pair
of hyperboloids $x\cdot x = 1$ and c) the null plane $x^0 - x^1 =
0$. This leads to Dirac's ``three forms of dynamics''.

At the first sight, it seems that perennials \textit{are}
observables. Thus, $p_k$ are the momenta and $J_k$ are the angular
momenta of the particle. Such an association is, however, more
difficult for $p_0$ and $K_k$. It is clearly not reasonable to put
$-p_0$ equal to the energy, because only $\sqrt{{\mathbf{p}}^2 + m^2}$
will lead to the observed spectrum of the particle. Of course,
$\sqrt{{\mathbf{p}}^2 + m^2}$ is the value of $-p_0$ at the constraint
surface $\Gamma$. More problems are encountered, if we try to
interpret $K_k$ as an observable. We can utilize $p_\mu$ together with
$K_k$ to form the functions
\[
  X_t^k := \frac{K_k}{p_0} + \frac{p_k}{p_0}t
\]
for any constant $t$: as functions of perennials, $X_t^k$ are
themselves perennials. They can be interpreted as coordinates of the
particle at the time $x^0 = t$. It turns out, that this is quite
general way of forming observables from perennials; for it to work,
some ``time structure'' is necessary. Here, we have used the
(standard) family of transversal surfaces $\Gamma_t$ defined by the
equations $x^0 = t$, as well as the family of maps generated by $p_0$
and sending $\Gamma_t$ to $\Gamma_{t'}$ for each pair $(t,t')$. One
easily verifies that
\[
  \frac{\partial X_t^k}{\partial t} + \{X_t^k,p_0\} = 0.
\]
The class $\{X_t^k\,|\,t \in {\mathbf{R}}\}$ can, therefore, be
considered as a Heisenberg observable, and so an observable turns out
to be a class of perennials.  The apparently trivial relation between
perennials and observables for the particular perennials $p_k$ and
$J_k$ in this case (the functions in the class are $t$-independent and
coincide in form with the original perennials) is due to the fact that
these perennials are connected in a special way to the standard time
structure (represented by $\Gamma_t$ and $p_0$) that is determined in
the phase space by an inertial frame: $p_k$ and $J_k$ have vanishing
Poisson brackets with $p_0$ and they leave (via Poisson bracket) each
time instant $\Gamma_t$ invariant. A systematic theory is given in
\cite{PH5}.

An important technical tool for the present paper will be
\textit{projection to a transversal surface}. One can project
perennials and symmetries. The projection $o_1$ of a perennial $o$ to
the transversal surface $\Gamma_1$ is a function on $\Gamma_1$ defined
as the pull-back of $o$ to $\Gamma_1$: $o_1 := o|_{\Gamma_1}$. This
projection preserves three operations: linear combinations,
multiplications of functions and Poisson brackets (of perennials). In
particular, for the Poisson brackets, we have: $\{o_1,o'_1\}_1 =
\{o,o'\}$, where $\{\cdot,\cdot\}_1$ is the Poisson bracket in
$(\Gamma_1,\Omega_1)$, $\{\cdot,\cdot\}$ that of
$(\tilde{\Gamma},\tilde{\Omega})$, $o_1$ the projection of $o$ and
$o'_1$ that of $o'$ to $\Gamma_1$ (for the proof, see
\cite{PH1}). Projections of perennials to a given transversal surface
can (sometimes!) be considered as Schr\"{o}dinger observables (for
more detail, see \cite{PH5}).

Similarly to perennials, symmetries can also be projected. Let $\phi$
be a symmetry and $p \in \Gamma_1$. Then, the projection $\phi_1 :
\Gamma_1 \rightarrow \Gamma_1$ of $\phi$ to $\Gamma_1$ is defined by
$\{\phi_1(p)\} := \gamma_{\phi(p)} \cap \Gamma_1$, where $\{a\}$ is
the set with the element $a$ and $\gamma_q$ is the c-orbit through the
point $q$. One can show that $\phi_1$ preserves $\Omega_1$ as well as
the group multiplication of symmetries; thus, the projection of a
group of symmetries is a group of symplectic maps
\cite{PH5}. Moreover, if $o$ generates a one-dimensional group of
symmetries $\phi_t$, then $o_1$ generates a one-dimensional group of
symplectic maps $\psi_t$ and it holds that $\psi_t = \phi_{t1}$ for
all $t$, where $\phi_{t1}$ is the projection of $\phi_t$ to $\Gamma_1$
\cite{PH1}.

Finally, let $\Gamma_1$ and $\Gamma_2$ be two transversal surfaces, $p
\in \Gamma_1$ and $\rho : \Gamma_1 \rightarrow \Gamma_2$ defined by
$\{\rho(p)\} := \gamma_p \cap \Gamma_2$. Then, the pull-back of
$\Omega_2$ by $\rho$ is $\Omega_1$ and the pull-back of $o_2$ is $o_1$
for any perennial $o$ with projection $o_1$ to $\Gamma_1$ and $o_2$ to
$\Gamma_2$. Thus, $\rho$ realizes the equivalence of all $\Gamma_i$'s
\cite{PH1}.

Another important notion that we can illustrate with the relativistic
particle model is that of \textit{relation}. The projections of the
ten perennials $p_\mu$, $J_k$ and $K_k$ to the transversal surface
$\Gamma_0$ are linearly independent. However, there must be four
functional relations between them, because $\Gamma_0$ is only
six-dimensional. There cannot be more independent relations, as the
functions form a complete system. The relations can be written as
\[
  {\mathbf{p}}^2 = -m^2,\quad \epsilon^{\mu\nu\rho\sigma}p_\nu
  J_{\rho\sigma} = 0,
\]
where $J_{\rho\sigma} := \epsilon_{\rho\sigma\mu\nu}x^\mu
p^\nu$. These relations have a close connection to the values of
Casimir elements for this particular representation of the
Poincar\'{e} group. Pohlmayer \cite{pohl} has discussed the general
case.

In the quantum version, we shall try to preserve these relations so
that they become similar relations between operators (that may be
deformed: some additional terms proportional to $\hbar$ may
appear). There also can be spectral conditions like $p_0 < 0$, that
should be satisfied in the quantum theory. A unitary representation of the
group $G$ that preserves the relations between its generators and that
satisfies the spectral condition is called \textit{physical
representation}.

\section{The homogeneous gauge}
\label{sec:hom}
In this section, we return to the 2+1 gravity and briefly summarize
our starting assumptions and equations.  More detail can be found in
\cite{M1}, \cite{M2}.  We shall consider only the metric-torus sector
of the 2+1 gravity system. In arbitrary coordinates $x^1$ and $x^2$ on
an arbitrary spacelike surfaces $t =$ const with the manifold
structure $\Sigma$, the metric of the spacetime $M$ has the form
\[
  ds^2 = - N^2dt^2 + g_{ab}(dx^a - N^adt)(dx^b - N^bdt)
\]
and the ADM action for the model reads
\[
  S = \int_{\mathbf R}dt\int_\Sigma d^2x\,\left(\pi^{ab}\frac{\partial
  g_{ab}}{\partial t} - {{\mathcal NH}} - {{\mathcal N}^a{\mathcal
  H}_a}\right), 
\]
where
\[
  {\mathcal H} = - \frac{1}{\sqrt{g}}[\pi^{ab}\pi_{ab} - (\pi^a_a)^2] +
  \sqrt{g}{\mathcal R},
\]
\[
  {\mathcal H}_a = -2\nabla_b\pi^b_a,
\]
$\nabla_a$ is the covariant derivative associated with the metric
$g_{ab}$ and $\mathcal R$ is the curvature scalar of the two-surfaces $t
=$ const. 

The analysis of the model simplifies enormously, if one can choose the
so-called ``homogeneous gauge''. It is the choice of coordinates such
that the fields $g_{ab}$ and $\pi^{ab}$ are independent of $x^1$ and
$x^2$ \cite{M-M}, \cite{M2}. A rigorous proof that such coordinates
exist for each classical solution of the model has not yet been
published. If each solution, however, admits at least one Cauchy
surface of constant mean curvature (CMC), which seems to be a very
plausible conjecture, then the existence can be shown \cite{M1}. The
existence of the CMC surface has been proved by L.~Andersson for
genera higher than 1 and there are some ideas even for genus 1
\cite{anderss}. We will assume, that there is such a Cauchy surface;
in any case, one can consider this as a part of the definition of the
system, and if the conjecture is invalid, then this system will not be
completely equivalent to the 2+1 gravity.

In the homogeneous gauge, the metric can be taken in the form \cite{M2}:
\be
  ds^2 = -N(t)^2dt^2 + e^{2\mu(t)}(dx^1)^2 + e^{2\nu(t)}(dx^2 +
  \beta(t)dx^1)^2.
\label{metric} 
\ee
A straightforward calculation then leads to the action 
\[
  S = \int dt\,(p_\mu\dot{\mu} + p_\nu\dot{\nu} + p_\beta\dot{\beta} -
  {{\mathcal NH}}),
\]
where
\[
  {\mathcal H} = \frac{1}{2}(e^{-\mu-\nu}p_\mu p_\nu -
  e^{\mu-3\nu}p_\beta^2). 
\]
Moncrief then performs two canonical transformations; the first is:
\[
  q^1 = \nu - \mu,\quad q^2 = \beta,\quad q^3 = \nu + \mu,
\]
\[
  p_\mu = -p_1 + p_3,\quad p_\nu = p_1 + p_3,\quad p_\beta = p_2,
\]
so that the super-Hamiltonian becomes
\[
  {\mathcal H} = \frac{1}{2}e^{-q^3}(p_3^2 - p_1^2 - e^{-2q^1}p_2^2).
\]
These coordinates are advantageous for visualisation of the geometry
of the system. The extended phase space is $\tilde{\Gamma} = T^*{\mathbf
R}^3$ with the canonical coordinates $q^1$, $q^2$, $q^3$, $p_1$, $p_2$
and $p_3$; the meaning of $q^3$ and $p_3$ is
\[
   q^3 = \log\sqrt{g},\quad p_3 = \frac{\sqrt{g}}{\mathcal N}\frac{\partial
         q^3}{\partial t}.
\]
The constraint surface $\Gamma$ is the light cone in the momentum space:
\[
  p_3^2 - p_1^2 - e^{-2q^1}p_2^2 = 0.
\]
The $p_3>0$ half of the cone represents expanding, the $p_3<0$  half
contracting, and the cusp $p_3 = 0$ represents the static tori---solutions
with higher symmetry. We shall adhere to the convention that the coordinate 
$t$ on the tori spacetimes is future oriented for ${\mathcal N} > 0$. 

The structure of the constraint surface $\Gamma$ and of the physical
phase space $\bar{\Gamma} = \Gamma/\gamma$ is spoilt by the points of
higher symmetry \cite{LM}. The c-orbits with $p_3 \neq 0$ are curves,
those with $p_3 = 0$ are just points. As the static solutions form a
set of measure zero, we can cut them away. Thus, we consider only that
part of the system that satisfies the condition
\[
  p_3 \neq 0.
\]
The new system will have a disconnected phase space $\tilde{\Gamma}' =
\tilde{\Gamma}'_+ \cup \tilde{\Gamma}'_-$, a constraint surface
$\Gamma' = \Gamma'_+ \cup \Gamma'_-$ and a physical phase space
$\bar{\Gamma}' = \bar{\Gamma}'_+ \cup \bar{\Gamma}'_-$, where
$\tilde{\Gamma}'_+ := \{x \in \tilde{\Gamma}\,|\,p_3 > 0\}$,
$\Gamma'_+ := \Gamma \cap \tilde{\Gamma}'_+$, and $\bar{\Gamma}'_+ :=
\Gamma'_+/\gamma$ (similarly for $p_3 < 0$).

After this truncation, Moncrief performes the second transformation:
\[
  T = \ln|p_3| - q^3, \quad p_T = -p_3,
\]
the other variables remaining the same; the super-Hamiltonian then reads
\be
  {\mathcal H} = - \frac{e^T}{2p_T}(p_T^2 - p_1^2 - e^{-2q^1}p_2^2).
\label{s-hamilt}
\ee
The meaning of the variable $T$ is given by
\[
  g^{ab}K_{ab} = \frac{1}{2{\mathcal N}}g^{ab} 
 \frac{\partial g_{ab}}{\partial t} = \epsilon e^T,
\]
where $\epsilon = \pm 1$ and the sign is determined by $\epsilon =\
$sign\,$p_3 = -$sign\,$p_T$. Thus, $\epsilon$ is just the sign of the CMC of
the surface $t=$ const.

Martin's \cite{mart} constants of motion (perennials) are in these
coordinates given by \cite{M2}: 
\bea 
C_1 & = & -\frac{\epsilon}{2}e^T\{[e^{-q^1}+(q^2)^2e^{q^1}](p_T+p_1) 
          - 2e^{-q^1}p_1 + 2q^2e^{-q^1}p_2\},
\label{C1} \\ 
C_2 & = & -\frac{\epsilon}{2}e^T[e^{q^1}(p_T+p_1)], 
\label{C2} \\ 
C_3 & = & -\frac{\epsilon}{2}e^T[q^2e^{q^1}(p_T+p_1) + e^{-q^1}p_2],
\label{C3} \\ 
C_4 & = & \frac{1}{2}\{[e^{-2q^1} - (q^2)^2]p_2 + 2q^2p_1\},
\label{C4} \\ 
C_5 & = & \frac{1}{2}p_2, 
\label{C5} \\ 
C_6 & = & p_1 - q^2p_2.
\label{C6}  
\eea 
They can be expressed as a kind of loop integrals by means of the
original fields $g_{ab}(x)$ and $\pi^{ab}(x)$ \cite{mart}. 

Moncrief observed that the Poisson brackets of the variables $P_\mu$ and
$J^\mu$ defined by
\be
  P_0 := \frac{1}{2}(C_1 + C_2),\quad P_1 := \frac{1}{2}(C_1 - C_2),\quad P_2
  := C_3, 
\label{CP}
\ee
and
\[
  J^0 := -C_4 + C_5,\quad J^1 := -C_4 - C_5,\quad J^2 = -C_6,
\]
form a Lie algebra isomorphic to iso(1,2): if we introduce the abbreviations
\[
  A := A^\mu P_\mu,\quad C := C_\mu J^{\mu},
\]
then
\be 
  \{A,A'\} = 0,\quad \{A,C\} = (\varepsilon^{\rho\mu\nu}A_\mu
  C_\nu)P_\rho,\quad \{C,C'\} =(\varepsilon_{\rho\mu\nu}C^\mu C^\nu)J^\rho.
\label{iso}
\ee
Here, we raise and lower the indices of $X^\mu$ and $P_\mu$ by the
Minkowskian three-metric diag$(-1,1,1)$, $\varepsilon^{\rho\mu\nu}$ and
$\varepsilon_{\rho\mu\nu}$ are the usual antisymmetric 
symbols ($\varepsilon_{\rho\mu\nu}$ is not $\varepsilon^{\rho\mu\nu}$
with lowered indices).

The formulas (\ref{C1})--(\ref{C6}) imply the following four equations:
\be
  C_1C_2 - C_3^2 = \frac{1}{4}e^{2T}(p_T^2 - p_1^2 - e^{-2q^1}p_2^2),
\label{C-H}
\ee
\[
  \{C_4,{\mathcal H}'\} = \{C_5,{\mathcal H}'\} = \{C_6,{\mathcal H}'\} = 0,
\]
where
\[
  {\mathcal H}' = p_T^2 - p_1^2 - e^{-2q^1}p_2^2.
\]
Thus, all $C$'s are perennials.  

There also are some discrete symmetries that originate from
non-uniqueness of the metric (\ref{metric}) for a given torus geometry
(class group transformations, see e.g. \cite{terras}). We shall not
discuss the question whether these transformations are to be
considered as symmetries or as gauge transformations. If
${\mathbf{X}}_1$ and ${\mathbf{X}}_2$ form a basis of a lattice in
${\mathbf E}^2$ defining the torus, then the metric has the form
\[
  g_{11} = {\mathbf{X}}_1\cdot {\mathbf{X}}_1,\quad g_{12} =
  {\mathbf{X}}_1\cdot  {\mathbf{X}}_2,\quad g_{22} = {\mathbf{X}}_2\cdot
  {\mathbf{X}}_2, 
\]
where ${\mathbf{a}}\cdot{\mathbf{b}}$ denotes the scalar product of
the vectors $\mathbf{a}$ and $\mathbf{b}$ in ${\mathbf{E}}^3$. The
following two transformations of the basis generate the whole group of
the discrete transformations (``large diffeomorphisms''):
\[
  {\mathbf{X}}'_1 = {\mathbf{X}}_2,\quad {\mathbf{X}}'_2 = {\mathbf{X}}_1,
\]
so that
\be
  g'_{11} = g_{22},\quad g'_{12} = g_{12},\quad g'_{22} = g_{11},
\label{classg1}
\ee
and
\[
  {\mathbf{X}}'_1 = {\mathbf{X}}_1,\quad {\mathbf{X}}'_2 =
  {\mathbf{X}}_1 + {\mathbf{X}}_2, 
\]
so that
\be
  g'_{11} = g_{11},\quad g'_{12} = g_{11}+g_{12},\quad g'_{22} =
  g_{11}+2g_{12}+g_{22}, 
\label{classg2}
\ee
Using Eq.\ (\ref{metric}), we obtain from Eq.\ (\ref{classg1}) for
$\mu$, $\nu$ 
and $\beta$:
\[
  e^{2\mu'} = \frac{e^{2(\mu+\nu)}}{e^{2\mu}+\beta^2e^{2\nu}},\quad
  e^{2\nu'} = e^{2\mu}+\beta^2e^{2\nu},
\]
\[
  \beta' = \beta\frac{e^{2\nu}}{e^{2\mu}+\beta^2e^{2\nu}}.
\]
The corresponding transformation of $q^1$, $q^2$ and $q^3$ is
\bea
  e^{q^{\prime 1}} & = & e^{q^1}[(q^2)^2 + e^{-2q^1}],
\label{classg11} \\
  q^{\prime 2} & = & \frac{q^2}{(q^2)^2 + e^{-2q^1}},
\label{classg12} \\
  q^{\prime 3} & = & q^3.
\label{classg13}
\eea
This gives for the momenta
\beann
  p'_1 & = & \frac{(q^2)^2 - e^{-2q^1}}{(q^2)^2 + e^{-2q^1}}p_1 +
             \frac{2q^2e^{-2q^1}}{(q^2)^2 + e^{-2q^1}}p_2, \\
  p'_2 & = & 2q^2p_1 - [(q^2)^2 - e^{-2q^1}]p_2.
\eeann
One can then easily verify that
\[
  p^{\prime 2}_1 + e^{-2q^{\prime 1}}p^{\prime 2}_2 = p^{2}_1 +
  e^{-2q^{1}}p^{2}_2,
\]
so that the super-Hamiltonian (\ref{s-hamilt}) is invariant.

The transformation (\ref{classg2}) gives
\[
  \mu' = \mu,\quad \nu' = \nu, \quad \beta' =1 + \beta.
\]
Thus,
\be
  q^{\prime 1} = q^1,\quad q^{\prime 2} = 1 + q^2,\quad q^{\prime 3} = q^3,
\label{classg21}
\ee
and 
\[
  p'_1 = p_1,\quad p'_2 = p_2,\quad p'_3 = p_3.
\]
Again, ${\mathcal H}$ is conserved.

The transformation (\ref{classg11})--(\ref{classg13}) implies
\[
  C'_1 = C_2,\quad C'_2 = C_1,\quad C'_3 = C_3,
\]
and (\ref{classg21}) implies
\[
  C'_1 = C_1 + C_2 + 2C_3,\quad C'_2 = C_2,\quad C'_3 = C_2 + C_3.
\]
These are both integral transformations with determinant 1 as one expects for
loop variables, if the loops are just permuted or linearly combined.

\section{Group comletion of the phase space}
\subsection{Completion by ISO(1,2)}
\label{sec:extISO}
In this section, we will investigate which transformations are generated by
the perennials $P_\mu$ and $J^\mu$ in the phase space of the system.
This task 
will be simplified, if we use coordinates that are adapted to
the perennials in the following sense.  $C_1$, $C_2$
and $C_3$ action via Poisson brackets in the phase space can be projected to
the configuration space spanned by $T$, $q^1$ and $q^2$, and the projections
are the vector fields $\hat{C}_1$, $\hat{C}_2$ and $\hat{C}_3$ given
by replacing 
$p_T$, $p_1$ and $p_2$ by $\partial/\partial T$, $\partial/\partial q^1$ and
$\partial/\partial q^2$ in the expressions (\ref{C1})--(\ref{C3}). The
perennials $C_1$, $C_2$ and $C_3$ have vanishing Poisson brackets with each
other, so $\hat{C}_1$, $\hat{C}_2$ and $\hat{C}_3$ will define a holonomous
frame; the corresponding coordinates are the desired ones.  Let us first
simplify these vectors by the transformation
\[
  u = T,\quad v = q^1 -T,\quad y = q^2,
\]
so that
\beann
 \hat{C}_1 & = & -\frac{\epsilon}{2}e^{-v}[(1 + y^2e^{2u+2v})\partial_u +
               2y\partial_y - 2\partial_v], \\
 \hat{C}_2 & = & -\frac{\epsilon}{2}e^{2u+v}\partial_u, \\
 \hat{C}_3 & = & -\frac{\epsilon}{2}e^{2u+v}(y\partial_u +
               e^{-2u-2v}\partial_y). 
\eeann
Now, we look for pairs of independent functions that are annihilated by each
of these
differential operators. The method of characteristics suggests that we study
integral curves of the vector fields. The integral curve of
$\hat{C}_1$ is 
defined by:
\[
  \dot{u} = A(1+y^2e^{2u+2v}),\quad
  \dot{y} = A(2y),\quad
  \dot{v} = A(-2),
\]
where $A = -\epsilon e^{-v}/2$. Thus,
\[
  \frac{1}{y}\dot{y} + \dot{v} = 0,
\]
\[
  -2e^{-2u-v}\dot{u} + 2ye^v\dot{y} + (-e^{-2u-v} + y^2e^v)\dot{v} = 0,
\]
and we have:
\[
  \hat{C}_1(ye^v) = 0,\quad \hat{C}_1(e^{-2u-v} + y^2e^v) = 0.
\]
The integral curve of $\hat{C}_1$ satisfies
\[
  \dot{u} = -\frac{1}{2}e^{2u+v},\quad
  \dot{y} = 0,\quad
  \dot{v} = 0,
\]
hence,
\[
  \hat{C}_2v = 0,\quad \hat{C}_2y = 0.
\]
Similarly for $\hat{C}_3$: 
\[
  \dot{u} = By,\quad
  \dot{y} = Be^{-2u-2v},\quad
  \dot{v} = 0,
\]
where $B = -(\epsilon/2)e^{2u+v}$. Thus,
\[
  -e^{-2u-2v}\dot{u} + y\dot{y} = 0,
\]
and
\[
  \hat{C}_3v = 0,\quad \hat{C}_3(e^{-2u-2v} + y^2) = 0.
\]
The pair of independent functions we have found for each vector field
determines all functions that are annihilated by the field. We can
easily find three independent functions such that each vector field
annihilates exactly two of them. The results can be summarized in the
following table.
\begin{center}
\begin{tabular}{l|lll}
 $\hat{C}_1$ & & $ye^v$, & $e^{-2u-v} + y^2e^v$, \\ \hline
 $\hat{C}_2$ & $v$, & $ye^v$, & \\ \hline
 $\hat{C}_3$ & $v$, & & $e^{-2u-v} + y^2e^v$.
\end{tabular}
\end{center}
Hence, the following transformation will simplify the vector fields:
\[
  \xi = e^{-2u-v} + y^2e^v,\quad \eta = v,\quad \zeta = ye^v.
\]
Indeed, we obtain that
\[
  \hat{C}_1 = \epsilon e^{-\eta}\partial_\eta,\quad \hat{C}_2 =
  \epsilon\partial_\xi,\quad 
  \hat{C}_3 = -\frac{\epsilon}{2}\partial_\zeta.
\]
Finally, the transformation
\[
  X^0 = \epsilon (e^\eta + \xi),\quad X^1 = \epsilon (e^\eta -
  \xi),\quad X^2 = 
  -2\epsilon\zeta, 
\]
gives
\[
  \hat{P}_\rho = \frac{\partial}{\partial X^\rho},\quad \rho = 0, 1, 2.
\]
Composing all the transformations, we express $X^\rho$ by means of the 
original variables $T$, $q^1$ and $q^2$:
\bea
 X^0 & = & \epsilon e^{-T}[e^{q^1} + e^{-q^1} + (q^2)^2e^{q^1}],
\label{X0} \\
 X^1 & = & \epsilon e^{-T}[e^{q^1} - e^{-q^1} - (q^2)^2e^{q^1}],
\label{X1} \\
 X^2 & = & \epsilon e^{-T}[-2q^2e^{q^1}].
\label{X2}
\eea

The class group transformations in terms of the coordinates $X$ read:
\beann
  X^{\prime 0} & = & (3/2)X^0 + (1/2)X^1 - X^2, \\
  X^{\prime 1} & = & -(1/2)X^0 + (1/2)X^1 + X^2, \\
  X^{\prime 2} & = & - X^0 - X^1 + X^2, 
\eeann
which is an element of SO(1,2), and 
\[ 
  X^{\prime 0} = X^0,\quad X^{\prime 1} = -X^1,\quad X^{\prime 2} = X^2.
\]

The relations (\ref{X0})--(\ref{X2}) together with
(\ref{C1})--(\ref{C3}) and (\ref{CP}) define a symplectic embedding
$\iota : \tilde{\Gamma}' \mapsto T^*{\mathbf M}^3$ of the phase space
$\tilde{\Gamma}'$ of our system, each component of which is spanned by
the coordinates $T$, $q^1$, $q^2$, $p_t$, $p_1$ and $p_2$, into the
cotangent bundle $T^*{\mathbf M}^3$ of the three-dimensional Minkowski
space ${\mathbf M}^3$ with the coordinates $(X^\mu,P_\mu)$. A very
important point is that the image $\iota(\tilde{\Gamma}')$ is only a
proper subset of $T^*{\mathbf M}^3$, namely the cotangent bundle of the
inside of the light cone of the origin. Indeed, calculating $X\cdot X$
from (\ref{X0})--(\ref{X2}), we obtain the identity
\[
 X\cdot X = -4e^{-2T} = -\frac{4}{\tau^2}.
\]
At the points of the light cone, the CMC is infinite; this surface
represents the singularity of the torus dynamics. For the CMC $\tau$,
we obtain 
\be 
  \tau = \frac{2\epsilon}{\sqrt{-X\cdot X}}.
\label{tau}
\ee

From Eq.\ (\ref{X0}), it follows that $\epsilon X^0>0$. Thus
$\tilde{\Gamma}'_+$ ($\tilde{\Gamma}'_-$) is mapped on the inside of
the future (past) light cone. Moreover, Eqs. (\ref{C1})--(\ref{C3})
and (\ref{CP}) yield
\[
  P_0 = \frac{\epsilon e^T}{4}(V_Tp_T - V_1p_1 - e^{-2q^1}V_2p_2),
\]
where
\[
 V_T = e^{q^1} + e^{-q^1} + (q^2)^2e^{q^1},\quad 
 V_1 = e^{q^1} - e^{-q^1} + (q^2)^2e^{q^1},\quad V_2 = 2q^2e^{q^1}.
\]
we easily verify the identity:
\[
 -V_T^2 + V_1^2 + e^{-2q^1}V_2^2 = -4.
\]
Thus, $(V_T,V_1,V_2)$ is a ``timelike vector'' oriented towards future
($V_T > 0$) and $(p_T,p_1,p_2)$ is a ``null vector'' (at the constraint
surface, see (\ref{s-hamilt})). Their ``scalar product'' $-V_Tp_T +
V_1p_1 + e^{-2q^1}V_2p_2$ is, therefore, negative (positive) if $p_T >
0$ ($p_T < 0$). As the sign of $p_T$ and of $\epsilon$ are correlated,
it follows that
\[
  P_0 < 0
\]
everywhere at the constraint surface. This, together with the Eqs.\
(\ref{CP}) and (\ref{C-H}) imply that the points of the constraint
surface satisfy the conditions
\be
  P\cdot P = 0,\quad P_0 < 0
\label{constrP}
\ee
with respect to the new variables $(X^\mu,P_\mu)$. Let us denote by
$\mathcal P$ the set of points in the momentum space with the coordinates
$P_\mu$ that satisfy Eq.\ (\ref{constrP}).

The transformation inverse to (\ref{X0})--(\ref{X2}) is well-defined
only inside the light cone, and can 
be written with respect to the coordinates as follows
\[
  e^T = \frac{2}{\sqrt{-X\cdot X}},\quad e^{q^1} = \epsilon
  \frac{X^0+X^1}{\sqrt{-X\cdot X}},\quad q^2 = -\frac{X^2}{X^0+X^1}.
\]
As $X^0 + X^1$ is positive (negative) inside the future (past) half
cone, $e^{q^1}$ will be always positive. If we try to extend the
transformation to the whole of ${\mathbf M}^3$, then we discover the
meaning of the points lying outside the light cone. A simple
calculation leads to the following complex transformation of the
original variables:
\[
  t = \mbox{i}t',\quad \nu = \nu' +\frac{\mbox{i}\pi}{2},
\] 
$\mu$ and $\beta$ remaining unchanged, where $t'$ and $\nu'$ are real.
Then, all new momenta are real, but
\[
  e^{2T} = - p_3^2e^{-2\mu-2\nu'}, \quad e^{2q^1} = - e^{-2\mu +
  2\nu'},
\]
so that $e^{2T}$ and $e^{2q^1}$ become negative as necessary. Thus, we
obtain spacetimes with the metric
\[
  ds^2 = {\mathcal N}^2(t')dt^{\prime 2} + e^{2\mu{t'}}(dx^1)^2 -
  e^{2\nu'(t')}[dx^2 + \beta(t')dx^1]^2,
\]
which have the Lorentz signature, but are acausal. A complete null
geodesic crossing from the inside to the outside of the light cone
represents an analytic three-dimensional spacetime analogous to the
Taub-NUT solution (see e.g.\ \cite{HE}) with a Cauchy horizon and the
Taub-NUT-like incompletenes at the cross point, if the singularity is
viewed as a lightlike torus.

Martin's perennials are push-forwarded by $\iota$ just into the usual
generators of the Poincar\'{e} group in the three-dimensional
Minkowski spacetime ${\mathbf M}^3$: $P_\rho$ and $J^\rho =
\varepsilon^{\rho\mu\nu}X_\mu P_\nu$. The corresponding group action
is not transitive on $T^*{\mathbf M}^3$: the orbits are classified by the
well-known invariants $P\cdot P$ and sign\,$P_0$, if $P\cdot P \leq
0$. However, each orbit of the group ISO(1,2) intersects
$\iota(\tilde{\Gamma}')$, so no superfluous orbit has been
added. Inside of the light cone, only the subgroup SO(1,2) acts; $P$'s
do not define any group acion on $\iota(\tilde{\Gamma}')$, because the
corresponding vector fields are badly incomplete there. We can
interpret our construction as follows.

The system of six Martin's functions form a complete algebra
perennials; only three of them, the generators of SO(1,2), can be
integrated to give a group action on the phase space
$\tilde{\Gamma}'$; the three Hamiltonian vector fields corresponding
to the perennials $P_\rho$ are incomplete in
$\tilde{\Gamma}'$. However, the Lie algebra generated by the six
perennials defines a group, ISO(1,2). With the standard symplectic
form, $\Omega_{\mbox{\footnotesize ISO}} = dP_\mu\wedge dX^\mu$ of
cotangent bundles, $T^*{\mathbf M}^3$ is a phase space, on which this
group does act. There is a map, $\iota$, that sends $\tilde{\Gamma}'$
in $T^*{\mathbf M}^3$; $\iota$ is a symplectic imbedding and it pushes
forwards Martin's perennials into the generators of the action of
ISO(1,2) on $T^*{\mathbf M}^3$. Thus, the map $\iota$ is equivariant for
the two actions of the subgroup SO(1,2) of ISO(1,2). Such a space
$T^*{\mathbf M}^3$ together with such a map $\iota$ can be called 
\textit{minimal group completion of the phase space $\tilde{\Gamma}'$
corresponding to the complete Lie algebra of Martin's perennials}. The
completion is minimal in the sense, that there is no smaller completion
(subspace of $T^*{\mathbf M}^3$), because each orbit of ISO(1,2) in
$T^*{\mathbf M}^3$ intersects $\iota(\tilde{\Gamma}')$.

The constraint surface $\Gamma'$ will be mapped to the subset of
$T^*{\mathbf M}^3$ given by $X\cdot X < 0$ and Eq.\ (\ref{constrP}). The
group ISO(1,2) does not act on $\iota(\Gamma')$ even if the generators
of the group are tangential to $\iota(\Gamma')$: again, the
translations are incomplete within this surface. However, there is a
unique completion of $\iota(\Gamma')$ in $T^*{\mathbf M}^3$ on which the
group acts, which we call $\Gamma_{\mbox{\footnotesize ISO}}$. This
surface is determined just by the equations (\ref{constrP}).  Clearly,
$(T^*{\mathbf M}^3, \Omega_{\mbox{\footnotesize ISO}},
\Gamma_{\mbox{\footnotesize ISO}})$ is a reparametrization invariant
system that defines the corresponding c-orbits: they coincide with the
maximal null geodesics in ${\mathbf M}^3$. Moreover, the image
$\iota(\gamma)$ of each c-orbit $\gamma$ in $\Gamma'$ lies completely
within some of the c-orbits of $(T^*{\mathbf M}^3,
\Omega_{\mbox{\footnotesize ISO}}, \Gamma_{\mbox{\footnotesize ISO}})$
namely in that null geodesic that extends $\iota(\gamma)$. Thus, we
also have a well-defined map, $\bar{\iota} : \bar{\Gamma}' \mapsto
\bar{\Gamma}_{\mbox{\footnotesize ISO}}$, where
$\bar{\Gamma}_{\mbox{\footnotesize ISO}}$ is the physical phase space
of $(T^*{\mathbf M}^3, \Omega_{\mbox{\footnotesize ISO}},
\Gamma_{\mbox{\footnotesize ISO}})$.
 
In this sense, we can speak about a \textit{group completion of the
reparametrization invariant system}.

The completion constructed above has an important property which makes
them interesting for physics. Let $(M,\Omega)$ be a symplectic
manifold and let a set ${\mathbf g}$ of functions form a Lie algebra with
respect to linear combinations and Poisson brackets. Let $G$ be the
(abstract) simply connected group that is determined by ${\mathbf g}$. Let
$(M_G,\Omega_G, \iota)$ be a minimal group completion of $(M,\Omega)$
by $G$. Thus, $G$ acts on $(M_G,\Omega_G)$ as a group of symplectic
diffeomorphisms. Let the image $\iota(M)$ be an open dense subset of
$M_G$. Then, we say that the group $G$ has a \textit{weak action} on
$(M,\Omega)$. For the above construction, we shall show the theorem:
\begin{thm}
The group ISO(1,2) has a weak action on the physical phase space
$\bar{\Gamma}'$. 
\end{thm}
{\textbf Proof} Consider the two surfaces $\Sigma_\pm$ defined by
$X\cdot X = 1$ and $\pm X^0 > 0$, respectively, inside the light cone
of the origin in ${\mathbf M}^3$; the manifolds $\Sigma_\pm\times
{\mathcal P}$ are global transversal surfaces in $\iota(\Gamma'_+)$
and $\iota(\Gamma'_-)$, as they are intersected by all null geodesics
inside of the light cone. Thus, $\bar{\Gamma}'$ can be identified with
$(\Sigma_+\times {\mathcal P}) \cup (\Sigma_-\times {\mathcal
P})$. Next consider the surface $\Sigma$ in ${\mathbf M}^3$ defined by
$X^0 = 0$. Clearly, the manifold $\Sigma\times {\mathcal P}$ is a
global transversal surface for the group completed system, because it
is intersected by any null geodesic at exactly one point; we can
identify $\bar{\Gamma}_{\mbox{\footnotesize ISO}}$ with $\Sigma\times
{\mathcal P}$.

Every point of $\Sigma_\pm\times {\mathcal P}$ defines a unique null
geodesic; this geodesic intersects $\Sigma\times {\mathcal P}$ at
precisely one point. Thus we have a well-defined map $\rho :
(\Sigma_+\times {\mathcal P}) \cup (\Sigma_-\times {\mathcal P}) \mapsto
\Sigma\times {\mathcal P}$ (the map $\rho$ and its properties have been
described in Sec.\ \ref{sec:particle}). It is easy to see that
$\bar{\iota}$ can be identified with $\rho$ and so we have to show
that $\rho((\Sigma_+\times {\mathcal P}) \cup (\Sigma_-\times {\mathcal P})$
is open and dense in $\Sigma\times {\mathcal P}$.

Let us introduce the coordinates $u$ and $v$ at the surfaces
$\Sigma_\epsilon$ 
by 
\beann X^0 & = & \epsilon\sqrt{1+u_1^2+u_2^2}, \\ 
X^1 & = & u_1, \\ 
X^2 & = & u_2, 
\eeann 
and the coordinates $x_1$ and $x_2$ at $\Sigma$ by $X^1 =
x_1$ and $X^2 = x_2$. Let us consider null geodesics with a definite
three-momenta of the form
\be 
P_\mu = p(1, \cos\alpha, \sin\alpha),
\label{mom}
\ee
where $p$ is a (negative) number. The null geodesic with the momenta
(\ref{mom}) starting at the point $(x_1,x_2)$ of $\Sigma$ will intersect
$\Sigma_\epsilon$ 
at the point
\beann
  t & = & \epsilon\sqrt{1+u_1^2+u_2^2}, \\
  u_1 & = & x_1 + \epsilon\cos\alpha\sqrt{1+u_1^2+u_2^2}, \\
  u_2 & = & x_2 + \epsilon\sin\alpha\sqrt{1+u_1^2+u_2^2}, 
\eeann
Solving for $(x_1,x_2)$, we obtain a description of $\rho$ in
terms of the coordinates $u_1$, $u_2$, $\epsilon$, $x_1$ and $x_2$:
\bea
  x_1 & = & u_1 - \epsilon\cos\alpha\sqrt{1+u_1^2+u_2^2},
\label{xu1} \\
  x_2 & = & u_2 - \epsilon\sin\alpha\sqrt{1+u_1^2+u_2^2}.
\label{xu2} 
\eea
To see which part of $\Sigma$ is hit, we introduce new variables:
\beann
  x'_1 & = & x_1\cos\alpha + x_2\sin\alpha, \\
  x'_2 & = & -x_1\sin\alpha + x_2\cos\alpha, \\
  u'_1 & = & u_1\cos\alpha + u_2\sin\alpha, \\
  u'_2 & = & -u_1\sin\alpha + u_2\cos\alpha, 
\eeann
and observe that
\[
  u_1^2+u_2^2 = u_1^{\prime 2}+u_2^{\prime 2}.
\]
Then, Eqs.\ (\ref{xu1}) and (\ref{xu2}) are equivalent to
\be
  x'_1 = u'_1 - \epsilon\sqrt{1+u_1^{\prime 2}+u_2^{\prime
  2}},\quad x'_2 = u'_2.
\label{xuprime}
\ee
Thus, as $u'_2$ runs through ${\mathbf R}$, so does $x'_2$. For
a fixed $x'_2$,
\[
  x'_1 = u'_1 - \epsilon\sqrt{1+u_1^{\prime 2}+x_2^{\prime
  2}},
\]
hence
\[
  \frac{dx'_1}{du'_1} = 1 - \epsilon\frac{u'_1}{\sqrt{1+u_1^{\prime
  2}+x_2^{\prime 2}}} > 0,  \epsilon = \pm 1. 
\]
For $\epsilon = +1$, $x'_1 \rightarrow -\infty$ as $u'_1
\rightarrow -\infty$. On the other side, we obtain
\[
  \lim_{u'_1=\infty}x'_1 = \lim_{u'_1=\infty}
  \frac{-1-x_2^{\prime 2}}{u'_1 + \sqrt{1+u_1^{\prime
  2}+x_2^{\prime 2}}} = 0.
\]
For $\epsilon = -1$, $x'_1 \rightarrow \infty$ as $u'_1
\rightarrow \infty$. On the other side, we obtain
  \[
  \lim_{u'_1=-\infty}x'_1 = \lim_{u'_1=-\infty}
  \frac{1+x_2^{\prime 2}}{|u'_1| + \sqrt{1+u_1^{\prime
  2}+x_2^{\prime 2}}} = 0.
\]
Thus, $\Sigma_+$ is mapped to $x_1\cos\alpha + x_2\sin\alpha < 0$ and
$\Sigma_-$ is mapped to $x_1\cos\alpha + x_2\sin\alpha > 0$. Only the
straight line $x_1\cos\alpha + x_2\sin\alpha = 0$ is missing from each
surface $P_1 =$ const, $P_2 =$ const. It follows that
$\rho(\tilde{\Gamma}')$ is open and dense, Q.E.D. 

There is a standard way of construction of group completions, if the
Lie algebra of observables is complete. Let $(M,\Omega)$ be a
symplectic manifold and let ${\mathbf g}$ be a Lie algebra generated
by a complete system of functions on $M$. Let $G$ be the unique simply
connected group whose Lie algebra coincides with ${\mathbf g}$. Let
${\mathbf g}^*$ be the dual linear space to ${\mathbf g}$ and let
ad$^*$ be the co-adjoint representation of $G$ on ${\mathbf g}^*$. The
orbits of the action ad$^*$ of $G$ in ${\mathbf g}^*$ are homogeneous
symplectic spaces of the group $G$ according to a beautiful result of
Kirillov \cite{kirill}. Moreover, from a basis of ${\mathbf g}$, a
(basis-independent) map $\Pi : M \mapsto \omega$ of $M$ into an orbit
$\omega$ can be constructed; $\Pi$ is the so-called momentum
map. Then, the manifold $\omega$ with Kirillov's symplectic structure
and with $\Pi$ as $\iota$ is the desired (minimal) completion.

As an example consider the manifold $M$ with the coordinates $q$ and
$p$ given by $q^2 + p^2 < 1$, equipped with the symplectic form
$\Omega = dp \wedge dq$. Let the algebra ${\mathbf g}$ be generated by
the functions $q$, $p$ and 1 (constant function). The corresponding
group is the three-dimensional Heisenberg group defined on ${\mathbf
R}^3$ with group law
\[
  (a_1,b_1,r_1)\cdot(a_2,b_2,r_2) = (a_1+a_2, b_1+b_2,
  r_1+r_2+\frac{b_1a_2-b_2a_1}{2}).
\]
The space dual to the algebra is ${\mathbf R}^3$ with the coordinates
$A$, $B$ and $R$, the orbits of the group are the planes $R =\ $const
and the momentum map is
\[
  A = q,\quad B = p,\quad R = 1.
\]
Thus, the orbit $\omega$ on which $M$ is mapped is given by the
equation $R = 1$. The image of $M$ is the disk $A^2 + B^2 < 1$, and
the group does not act even weakly.

\subsection{SO(2,3) completion}
\label{sec:SO(2,3)}
There is a motivation to look for further symmetries: the subgroup
SO(1,2) that acts on $\tilde{\Gamma}'$ is too small to define a time
evolution \`{a} la Dirac. The simple form of the reparametrization
invariant system $(T^*{\mathbf M}^3, \Omega_{\mbox{\footnotesize ISO}},
\Gamma_{\mbox{\footnotesize ISO}})$ allows us to see immediately that
there are more perennials than just the generators of the
three-dimensional Poincar\'{e} group: we have also the conformal
isometries. The so-called dilatation is generated by \be D := X^\mu
P_\mu,
\label{D}
\ee
and the so-called conformal accelerations are generated by
\bea
  Q_0 & := &  (X\cdot X) P_0 + 2X^0(X\cdot P), 
\label{Q0} \\
  Q_1 & := &  (X\cdot X) P_1 - 2X^1(X\cdot P),
\label{Q1} \\
  Q_2 & := &  (X\cdot X) P_2 - 2X^2(X\cdot P).
\label{Q2}
\eea
It is easy to verify that the Poisson brackets of these variables with
$P\cdot P$ weakly vanish. Let us denote $B^\mu Q_\mu$ by $B$. Then,
the Lie algebra of $A$, $B$, $C$ and $D$ is given by Eqs.\ (\ref{iso})
and 
\be
  \{A,B\} =  2(A\cdot B)D - 2(\varepsilon_{\rho\mu\nu}A^\mu
B^\nu)J^\rho,\quad \{B,C\}  =  (\varepsilon^{\rho\mu\nu}B_\mu C_\nu)Q_\rho,
\label{so1}
\ee
\be 
  \{A,D\}  =  -A,\quad
  \{B,B'\}  =  0,\quad
  \{B,D\}  =  B,\quad
  \{C,D\}  =  0.
\label{so2}
\ee
This is the Lie algebra of the group SO(2,3). Indeed, let 
$Z^0$, $Z^1$, $Z^3$, $W^0$ and $W^1$ be coordinates in ${\mathbf R}^5$
with the 
metric 
\be
  dS^2 = -(Z^0)^2 - (dW^0)^2 + (dZ^1)^2 + (dZ^2)^2 + (dW^1)^2;
\label{5-metric}
\ee
to obtain the algebra (\ref{iso}), (\ref{so1}) and (\ref{so2}), we have
just to identify:
\[
  J^\rho \mapsto \varepsilon^{\rho\mu\nu}Z_\mu\frac{\partial}{\partial
  Z^\nu},\quad 
  D \mapsto W_0\frac{\partial}{\partial W^1} -
  W_1\frac{\partial}{\partial W^0}, 
\]
\[
  P_\rho \mapsto \left(Z_\rho\frac{\partial}{\partial W^0} -
  W_0\frac{\partial}{\partial 
  Z^\rho}\right) - \left(Z_\rho\frac{\partial}{\partial W^1} -
  W_1\frac{\partial}{\partial 
  Z^\rho}\right),
\]
\[
  Q_\rho \mapsto \left(Z_\rho\frac{\partial}{\partial W^0} -
  W_0\frac{\partial}{\partial 
  Z^\rho}\right) + \left(Z_\rho\frac{\partial}{\partial W^1} -
  W_1\frac{\partial}{\partial 
  Z^\rho}\right),
\]
where the indices are lowered by the metric (\ref{5-metric}).

The Hamiltonian vector field corresponding to the dilatation is complete not
only within $T^*{\mathbf M}^3$, but even within
$\iota(\tilde{\Gamma}')$. The Hamiltonian vector fields corresponding
to $Q$'s are incomplete within $T^*{\mathbf M}^3$; however, there is still
a chance to construct additional observables from $Q$'s, so we have
to construct the next completion. This completion is well-known: it is the
cotangent bundle $T^*\bar{{\mathbf M}}^3$ of the compactified Minkowski
spacetime $\bar{{\mathbf M}}^3$ \cite{P-R}. Let us briefly describe the
construction, because we shall need some details of it.

Consider the three-dimensional Einstein cosmology  spacetime
${\mathbf M}_E^3$ with the coordinates $\tau$, $\vartheta$ and $\varphi$ and
the metric
\[
  ds^2 = - d\tau^2 + d\vartheta^2 + \sin^2\vartheta d\varphi^2.
\]
The null geogesics that start at the point $\tau = \tau_0$, $\vartheta = 0$,
are given by
\[
  \tau = \tau_0 + \lambda,\quad \vartheta = \lambda,\quad \varphi =
  \mbox{const}. 
\]
These geodesic form a null cone that refocuses at the point $\tau = \tau_0 +
\pi$, $\vartheta = \pi$. Similarly, all null geodesics through the
point $\tau 
= \tau_0$, $\vartheta = \pi$, have the form
\[
  \tau = \tau_0 + \lambda,\quad \vartheta = \pi - \lambda,\quad \varphi =
  \mbox{const}, 
\]
and they refocuse at $\tau = \tau_0 + \pi$, $\vartheta = 0$. As it is
well-known, (e.g.\ \cite{P-R}), the compactified Minkowski spacetime
$\bar{{\mathbf M}}^3$ is obtained from ${\mathbf M}_E^3$ by the
identification of each point $(\tau,\vartheta,\varphi)$ with the point
$(\tau + \pi,\pi - \vartheta,\varphi + \pi)$. There is a conformal
isometry $\phi$ that sends ${\mathbf M}^3$ into $\bar{{\mathbf M}}^3$
such that the set $\phi({\mathbf M}^3)$ lies in the future of the null
cone between the points $(\tau = - \pi, \vartheta = 0)$ and $(\tau =
0, \vartheta = \pi)$, and in the past of the null cone between $(\tau
= 0, \vartheta = \pi)$ and $(\tau = \pi, \vartheta = 0)$.

Another copy of Minkowski spacetime lies beween the points $(\tau = -\pi,
\vartheta = \pi)$, $(\tau = 0, \vartheta = 0)$ and $(\tau = \pi, \vartheta =
\pi)$. By the above identification and the two conformal isometries,
the point 
$X^\mu$ of the first Minkowski spacetime will be mapped to the point $Y^\mu$
of the second one given by 
\be
  Y^\mu = \frac{X^\mu}{X\cdot X},
\label{inverse}
\ee
if the inertial coordinates $X^\mu$ and $Y^\mu$ are chosen properly. We will
make some use of these two patches of $\bar{{\mathbf M}}^3$; the fact that
they do not cover $\bar{{\mathbf M}}^3$ will not be important. Let us
call them $U$ and $V$.

The map $\phi$ is a diffeomorphism of three-dimensional manifolds and it can
be extended to an isomorphism $\phi_{\mbox{\footnotesize{cot}}}$ of the
cotangent bundles of these manifolds. As the conformal group SO(2,3) acts
transitively in $T^*\bar{{\mathbf M}}^3$, the triad $(T^*\bar{{\mathbf M}}^3,
\Omega_{\mbox{\footnotesize SO}}, \phi_{\mbox{\footnotesize{cot}}})$
is the desired minimal SO(2,3) completion; the form
$\Omega_{\mbox{\footnotesize SO}}$ is the standard symplectic form of
cotangent bundles; observe that it is exact. The set
$\phi_{\mbox{\footnotesize{cot}}}(T^*{\mathbf M}^3)$ is equal to $T^*U$,
and so it is open and dense in $T^*\bar{{\mathbf M}}^3$. Thus, SO(2,3)
\textit{acts weakly} on $T^*{\mathbf M}^3$.

Let the canonical coordinates in the cotangent bundles $T^*U$ and $T^*V$ be
$X^\mu$, $P_\mu$ and $Y^\mu$, $Q_\mu$, respectively. Then the transformation
(\ref{inverse}) between $X^\mu$ and $Y^\mu$ leads to the following
transformation between $P_\mu$ and $Q_\mu$:
\be
  Q_\mu = (X\cdot X) P_\mu - 2(X\cdot P)X_\mu.
\label{Qmu}
\ee
The inverse map is
\be
  X^\mu = \frac{Y^\mu}{Y\cdot Y},\quad P_\mu = (X\cdot X) Q_\mu - 2(X\cdot
  Q)Y_\mu.
\label{Pmu}
\ee
Comparison with Eqs.\ (\ref{Q0})--(\ref{Q2}) shows that the use of the
letter $Q$ for this coordinate will not lead to any confusion with the
notation for the generators of conformal accelerations.  From the
transformation formulas (\ref{Qmu}) and (\ref{Pmu}), we easily verify
that the symplectic form $\Omega_{\mbox{\footnotesize SO}}$ at the
points where the patches overlap satisfies 
\[
  dP_\mu \wedge dX^\mu = dQ_\mu \wedge dY^\mu.
\]
The SO(2,3) completion of the RIS $(T^*{\mathbf M}^3,
\Omega_{\mbox{\footnotesize{ISO}}}, \Gamma_{\mbox{\footnotesize{ISO}}})$ is
$(T^*\bar{{\mathbf M}}^3, \Omega_{\mbox{\footnotesize{SO}}},
\Gamma_{\mbox{\footnotesize{SO}}})$, where 
$\Gamma_{\mbox{\footnotesize{SO}}}$ is given by the equations
\[
  g_i^{\mu\nu}P^i_\mu P^i_\nu = 0,\quad P^i_0 < 0
\]
within each chart $T^*U_i$, where the metric $g_i^{\mu\nu}$ is the
metric of the conformal chart $U_i$, $P^i_\mu$ the canonical coordinate in
the fibers of $T^*U_i$ and we allow only charts that have the same time
orientation. 

It follows that the projection of the c-orbits to the configuration
space $T^*\bar{{\mathbf M}}^3$ are complete null geodesics; they are
closed curves (topologically $S^1$). The $\phi$-images of the c-orbits
of the system $(T^*{\mathbf M}^3, \Omega_{\mbox{\footnotesize{ISO}}},
\Gamma_{\mbox{\footnotesize{ISO}}})$ are the null geodesics in the
chart $U$; each of them is completed by one point in $\bar{{\mathbf
M}}^3$. The null geodesics in $\bar{{\mathbf M}}^3$ that do not contain
any $\phi$-images form the light cone of the origin of the chart $V$
($Y^\mu = 0$). Hence, the surface $\Gamma_0$ given by the equations
$X^0 = 0$, $P_0 = - \sqrt{P_1^2 + P_2^2}$ within the chart $T^*U$, and
by $Y^0 = 0$, $Q_0 = - \sqrt{Q_1^2 + Q_2^2}$ within the chart $T^*V$
is a global transversal surface. We can introduce coordinates
$(x_1,x_2) \in {\mathbf R}^2$, $(p_1,p_2) \in {\mathbf R}^2 \setminus \{0\}$
in $\Gamma_0 \cap T^*U$ and $(y_1,y_2) \in {\mathbf R}^2$, $(q_1,q_2) \in
{\mathbf R}^2 \setminus \{0\}$ in $\Gamma_0 \cap T^*V$ such that the
imbedding equations are \be \left.
\begin{array}{lll}
  X^0=0,& X^1=x_1,& X^2=x_2, \\
  P_0=-\sqrt{p_1^2 + p_2^2},& P_1=p_1,& P_2=p_2,
\end{array}
\right\}
\label{coord}
\ee
and
\[
\begin{array}{lll}
  Y^0=0,& Y^1=y_1,& Y^2=y_2, \\
  Q_0=-\sqrt{q_1^2 + q_2^2},& Q_1=q_1,& Q_2=q_2.
\end{array}
\] 
One easily verifies that $\Gamma_0$ defined in
this way is a smooth surface and that the transformation formulas
(\ref{inverse}), (\ref{Qmu}) and (\ref{Pmu}) imply the relations
\be
  x_k = \frac{y_k}{y_1^2+y_2^2},\quad p_k = (y_1^2+y_2^2)q_k -
  2(y_1q_1+y_2q_2)y_k. 
\label{xy}
\ee
In particular, Eq.\ (\ref{Qmu}) implies that $Q_0 < 0$ if $P_0 < 0$. The
pull-back $\Omega_0$ of the symplectic form
$\Omega_{\mbox{\footnotesize{SO}}}$ to $\Gamma_0$ is
given in these coordinates by
\[
  \Omega_0 = dp_k \wedge dx_k = dq_k \wedge dy_k,
\]
the last inequality following from Eqs.\ (\ref{xy}). $\Omega_0$ is exact.

The manifold $\Gamma_0$ is a bundle with the fiber ${\mathcal P} \cong
S^1\times {\mathbf R}$ given by $x_k =$ const or $y_k =$ const. The base
space is $S^2$, and the coordinates $(x_1,x_2)$ and $(y_1,y_2)$ are
nothing but the two stereographic projection charts of $S^2$. The
symplectic space $(\Gamma_0, \Omega_0)$ is a homogeneous symplectic
space of the group SO(2,3), which acts on $\Gamma_0$ by projection of
symmetries (see Sec.\ \ref{sec:particle}): each element of the
conformal group maps null geodesics in null geodesics. $(\Gamma_0,
\Omega_0)$ can be identified with the group completed physical phase
space $\bar{\Gamma}_{\mbox{\footnotesize{SO}}}$. The composition
$\bar{\phi} \circ \bar{\iota}$ of the ISO(1,2)-completion and the
SO(2,3)-completion gives the image $\bar{\phi}(
\bar{\iota}(\bar{\Gamma}'))$ as an open dense subspace of $\Gamma_0$;
thus, the conformal group SO(2,3) acts weakly (and transitively) on
$\bar{\Gamma}'$.
 
The generators of the action of SO(2,3) on $\Gamma_0$ are Hamiltonian
vector fields of the projections of the perennials $P_\mu$, $Q_\mu$,
$J^\mu$ and $D$ from $T^*{\mathbf M}^3$ to $\Gamma_0$ (see Sec.\
\ref{sec:particle}). In the patch $(x_k,p_k)$, the projections
coincide with the functions \be \left.
\begin{array}{llll}
 \bar{P}_0 := -\sqrt{{\mathbf p}\cdot{\mathbf p}},&\bar{Q}_0 :=
 -({\mathbf x}\cdot{\mathbf x})\sqrt{{\mathbf p}\cdot{\mathbf p}},  &
  \bar{J}^0  :=
 x_1p_2 - x_2p_1,  \\  
 \bar{P}_1 := p_1, &
 \bar{Q}_1 := ({\mathbf x}\cdot{\mathbf x})p_1 - 
 2({\mathbf x}\cdot {\mathbf p})x_1, & \bar{J}^1 :=
 - x_2\sqrt{{\mathbf p}\cdot{\mathbf p}}, \\
 \bar{P}_2 :=
 p_2, & \bar{Q}_2 :=
 ({\mathbf x}\cdot{\mathbf x})p_2 - 
 2({\mathbf x}\cdot {\mathbf p})x_2, & \bar{J}^2 
 := x_1\sqrt{{\mathbf p}\cdot{\mathbf p}}, \\
 &&\bar{D} := {\mathbf x}\cdot {\mathbf p}.  
\end{array}
\right\}
\label{baralg}
\ee These functions will be considered as observables. Here, we have
used the abbreviation ${\mathbf u}\cdot{\mathbf v} := u_1v_1 + u_2v_2$.  The
expressions within the other patch, $(y_k,q_k)$, are analogous, one
just have to exchange $P$'s and $Q$', write $y$ for $x$ and $q$ for
$p$. Via Poisson brackets, the functions generate the Lie algebra of
SO(2,3). They are ten functions of four variables; thus, there will be
six relations. These relations can be systematically written down, if
we solve the definitions of $\bar{P}_1$, $\bar{P}^2$, $\bar{J}^1$, and
$\bar{J}^2$ for $p_1$, $p_2$, $x_1$ and $x_2$ and substitute the
results into the other definitions: 
\be 
\bar{P}_0^2 = \bar{P}_1^2 +
\bar{P}_2^2,\quad \bar{J}^0\bar{P}_0 + \bar{J}^1\bar{P}_1 +
\bar{J}^2\bar{P}_2 = 0,
\label{rel1}
\ee
\be
\quad \bar{Q}_0^2 = \bar{Q}_1^2 + \bar{Q}_2^2,\quad \bar{D} = 
\frac{\bar{P}_1\bar{J}^2 -
\bar{P}_2\bar{J}^1}{\sqrt{{\mathbf \bar{P}\cdot{\bar{P}}}}}, 
\label{rel2}
\ee
\be
\bar{Q}_1 = \frac{\bar{P}_1}{{\mathbf \bar{P}\cdot\bar{P}}}[(\bar{J}^1)^2 -
(\bar{J}^2)^2] +
\frac{\bar{P}_2}{{\mathbf \bar{P}\cdot{\bar{P}}}}[2\bar{J}^1\bar{J}^2], 
\label{rel3}
\ee
\be
\bar{Q}_2 =
\frac{\bar{P}_1}{{\mathbf \bar{P}\cdot\bar{P}}}[2\bar{J}^1\bar{J}^2] - 
\frac{\bar{P}_2}{{\mathbf \bar{P}\cdot\bar{P}}}[(\bar{J}^1)^2 -
(\bar{J}^2)^2]. 
\label{rel4}
\ee
The two Eqs.\ (\ref{rel1}) are relations concerning also ISO(1,2) alone and
the four relations (\ref{rel2})--(\ref{rel4}) can be used to calculate
the remaining generators of SO(2,3). A quadratic relation follows
\be
  - (\bar{J}^0)^2 + (\bar{J}^1)^2 + (\bar{J}^2)^2 = \bar{D}^2.
\label{quadrel}
\ee
The Poisson brackets of the four quadratic expressions $P\cdot P$,
$Q\cdot Q$, $J\cdot P$ and $-D^2 + J\cdot J$ with the generators
$P_\mu$, $Q_\mu$, $J^\mu$ and $D$ are mostly vanishing or proportional
to $P\cdot P$; they are (generalized) Casimirs of some subgroups.

As it was explained in Sec.\ \ref{sec:particle}, the action of SO(2,3)
on $\Gamma_0$ is generated by the Hamiltonian vector fields of the
observables (\ref{baralg}). This defines a linear map from the Lie
algebra so(2,3) into vector fields on $\Gamma_0$. We can describe the
map, if we choose a basis of so(2,3) and list the images of the
elements of the basis. Let the basis be 
\be 
({\mathcal P}_0,{\mathcal
P}_1,{\mathcal P}_2,{\mathcal Q}_0,{\mathcal Q}_1, {\mathcal
Q}_2,{\mathcal J}^0,{\mathcal 
J}^1, {\mathcal J}^2,{\mathcal D}).
\label{a-basis}
\ee 
Here, we denote the abstract elements of the Lie algebra by upper case
calligraphic letters to distinguish them from the corresponding perennials or
vector fields. The association with the vector fields is:
\be
 {\mathcal P}_0 \mapsto 
 -\frac{p_k}{\sqrt{{\mathbf p}\cdot{\mathbf p}}}\frac{\partial}{\partial
 x_k},\quad {\mathcal P}_k \mapsto \frac{\partial}{\partial x_k},
\label{hamvecP}
\ee
\be
 {\mathcal Q}_0 \mapsto 
 -\frac{{\mathbf x}\cdot{\mathbf x}}{\sqrt{{\mathbf p}\cdot{\mathbf p}}}\,p_k
 \frac{\partial}{\partial x_k} + 2\sqrt{{\mathbf p}\cdot{\mathbf p}}\,x_k
 \frac{\partial}{\partial p_k},
\label{hamvecQ0}
\ee
\be
 {\mathcal Q}_k \mapsto [({\mathbf x}\cdot{\mathbf x})\delta_{kl} - 2x_kx_l]
 \frac{\partial}{\partial x_l} + [({\mathbf x}\cdot{\mathbf p})\delta_{kl} +
 x_kp_l - x_lp_k]\frac{\partial}{\partial p_l},
\label{hamvecQk}
\ee
\be
 {\mathcal J}^0 \mapsto -x_2\frac{\partial}{\partial x_1} +
 x_1\frac{\partial}{\partial x_2} - p_2\frac{\partial}{\partial p_1} +
 p_1\frac{\partial}{\partial p_2},
\label{hamvecJ0}
\ee
\be
 {\mathcal J}^1 \mapsto 
 -\frac{x_2p_k}{\sqrt{{\mathbf p}\cdot{\mathbf
  p}}}\frac{\partial}{\partial x_k} 
 + \sqrt{{\mathbf p}\cdot{\mathbf p}}\frac{\partial}{\partial p_2},
\label{hamvecJ1}
\ee
\be
 {\mathcal J}^2 \mapsto 
 \frac{x_1p_k}{\sqrt{{\mathbf p}\cdot{\mathbf p}}}
 \frac{\partial}{\partial x_k}   
 - \sqrt{{\mathbf p}\cdot{\mathbf p}}\frac{\partial}{\partial p_1},
\label{hamvecJ2}
\ee
\be
 {\mathcal D} \mapsto x_1\frac{\partial}{\partial x_1} +
 x_2\frac{\partial}{\partial x_2} - p_1\frac{\partial}{\partial p_1}  
 - p_2\frac{\partial}{\partial p_2}.
\label{hamvecD}
\ee
We shall need the form of these vector fields for the construction of
the physical representation of the group in Sec.\ \ref{sec:repre}.

\section{The Hamiltonian}
\label{sec:hamilton}
In this section, we study the time evolution of the 2+1 gravity
model. We are going to apply Dirac's idea: a choice of transversal
surfaces of maximal symmetry, and a comparison of different time
levels using symmetry operations.

There are two problems that prevent a straightforward
application. First, the group SO(2,3) is too large to have a
representation that satisfies all conditions on \textit{physical}
representation (see the next section).  Second, only a
four-dimensional subgroup of SO(2,3) has the \textit{action} on the phase
space of the system that is associated with the corresponding Lie
algebra of perennials.

The largest subgroups that have physical representations are $G_1$ and
$G_2$ with the structure $(SO(1,2)\times {\mathbf R}) \otimes_S
{\mathbf R}^3$.  $G_1$ is generated by ${\mathcal J}^0$, ${\mathcal
J}^1$, ${\mathcal J}^2$, ${\mathcal D}$, ${\mathcal P}_0$, ${\mathcal
P}_1$ and ${\mathcal P}_2$ and $G_2$ by ${\mathcal J}^0$, ${\mathcal
J}^1$, ${\mathcal J}^2$, ${\mathcal D}$, ${\mathcal Q}_0$, ${\mathcal
Q}_1$ and ${\mathcal Q}_2$. Consider $G_1$. The corresponding group
completion of the 2+1 gravity coincides with the ISO(1,2)-completion
that was constructed in the previous section, because the only
additional element, the dilatation ${\mathcal D}$, acts on
$\tilde{\Gamma'}$. Let us, therefore, restrict ourselves to $G_1$.

The group $G_1$ acts weakly on the physical phase space
$\bar{\Gamma}'$. This is important for the quantum generators of the
group to have suitable spectra. On the other hand, the weak action is
not sufficient for the construction of a time evolution according to
Dirac. Let us explain these two points.

Consider the two-dimensional disk example described at the end of the
previous section. The Heisenberg group did not act even weakly. On the
disk, the values of the classical observables $q$ and $p$ are bounded:
$q^2 + p^2 < 1$. On the group completion, which is a whole plain, the
values of $q$ fill up the interval $(-\infty,\infty)$ and similarly
for $p$. This follows from the structure of the Lie algebra (see,
e.g. \cite{ishamG}). If we represent $q$ and $p$ by self-adjoint
operators satisfying the canonical commutation relations, then this
structure forces the spectra again to fill up the whole real axis. The
corresponding quantum mechanics contains semiclassical wave packets
with average values of $\hat{q}$ and $\hat{p}$ that lie far away from
possible classical values. Let us call this the \textit{problem of
ranges}. On the other hand, if a group acts weakly,
then the only change of the range of classical values that is
motivated is an addition of some limit points. This would happen in any
case in the quantum mechanics, because the spectrum of any
self-adjoint operator is a closed subset of ${\mathbf R}$. In this
respect, the weak action does not differ from an ordinary action.

However, for Dirac's idea to work, we have to find a family of
maximally symmetric transversal surfaces and a sufficiently large
subset of the symmetry group so that all such surfaces can be obtained
from one by the action of the subset. We emphasize that this has to work
within the constraint surface of the classical theory so that we can
interpret the transformations. Clearly, for these
purposes, the weak action is not adequate. First, the maximally
symmetric surfaces will lie in the group completed constraint surface, but
not, in general, inside that of the system; the intersection of such
surfaces with the constraint surface of the system will not be, in general,
globally transversal. Second, an image of a globally transversal
surface by an element of the group that has only a weak action will
not, in general, lie inside the constraint surface. It is easy to construct
examples of this kind for the action of $G_1$. This means that only
the subgroup $G_0$ is at our disposal for Dirac's construction.

The group $G_0 =$ SO(1,2)$\times {\mathbf R}$ generated by ${\mathcal
J}^0$, ${\mathcal J}^1$, ${\mathcal J}^2$ and ${\mathcal D}$ is a
cartesian product of two simple groups. Its three-dimensional
subgroups are of two types: 
\ben
\item
SO(1,2), which is the \textit{unique}
subgroup of this type, because it is a normal subgroup,
\item
subgroups that leave a null plane invariant; an example is the subgroup
of the plane $X^0 + X^1 = 0$, which is generated by ${\mathcal J}^0 +
{\mathcal J}^1$, ${\mathcal J}^2$ and ${\mathcal D}$. All other subgroups
of this 
type are (group-) similar to this one.
\een

The surfaces in $\Gamma'$ symmetric with respect to SO(1,2) satisfy
$-(X^0)^2 + (X^1)^2 +(X^2)^2 =$ const (const $< 0$) have two
components (with $X^0 > 0$ and $X^0 < 0$) and form one-dimensional
family. The components coincide with the CMC surfaces, and the union
of both components is globally transversal. On the other hand, the
projection of the surfaces to ${\mathbf M}^3$ that are invariant with
respect to the null plane groups are of course these null planes; the
surfaces do not lie inside the null cone $\iota(\tilde{\Gamma}')$ and
their intersections with $\iota(\tilde{\Gamma}')$ are not
transversal. We summarize the results:
\begin{thm}
The 2+1 gravity model possesses a unique one-dimensional family of maximally
symmetric globally transversal surfaces. Each such surface has two
components that are CMC surfaces with opposite values of CMC.
\end{thm}

The second step of the construction is to find a subgroup that would
carry us along the family of the CMC surfaces. Thus, it must be a
subgroup whose elements are representatives of all classes of
$G_0/$SO(1,2). However, $G_0/$SO(1,2) $= {\mathbf R}$, so the desired
subgroup is generated by just one element, which must have the form
${\mathcal D} + a{\mathcal J}^0 + b{\mathcal J}^1 + c{\mathcal J}^2$,
where $a$, $b$ and $c$ are three arbitrary reals. We can normalize the
generator in this way, as the overal factor does not change the
subgroup, and the factor in front of ${\mathcal D}$ must be non-zero:
a nontrivial motion of the CMC surface is generated just by the
${\mathcal D}$ term. Let us study how it acts on the CMC $\tau$. Using
equations (\ref{tau}) and (\ref{D}), we find that
\[
  \{\tau, D\} = - \frac{2\epsilon}{\sqrt{-X\cdot X}}.
\]
Thus, for $\epsilon > 0$, the action of $D$ diminishes $\tau$, and for
$\epsilon < 0$, it enlarges $\tau$. $\epsilon > 0$ ($\epsilon < 0$)
means that we are in the future (past) light cone of the origin. In
the future light cone, we have expanding tori ($\tau > 0$) and they
expand from the ``big bang'' $\tau = \infty$ to the maximal expansion
state $\tau = 0$. Thus, $D$ generates evolution towards future
here. In the past light cone ($\tau < 0$), we have contracting tori
that start at the maximal expansion state $\tau = 0$ and finish at the
``big crunch'' $\tau = -\infty$. Thus $D$ generates an evolution
towards past. Luckily enough, there is a smooth Hamiltonian $H$ that
evolves everything towards the future: $H = D$ inside the future light
cone and $H = -D$ inside the past light cone. Les us calculate the
corresponding perennial $\bar{H}$ on the physical phase space
$\bar{\Gamma}_{\mbox{\footnotesize ISO}}$. From the proof of the
theorem 1, it follows that the portion $\bar{\Gamma}'_+$ of the
physical phase space that corresponds to the future half of the light
cone is given by $x_1\cos\alpha + x_2\sin\alpha < 0$ and that
$\bar{\Gamma}'_-$ corresponding the past one by $x_1\cos\alpha +
x_2\sin\alpha > 0$. However, Eq.\ (\ref{mom}) implies that
\[
  p_1 = p\cos\alpha,\quad p_2 = p\sin\alpha,
\]
where $p < 0$. Here $x_1$, $x_2$, $p_1$ and $p_2$ are the coordinates that we
have chosen in the physical phase space $\bar{\Gamma}$ (cf.\ Eq.\
(\ref{coord})). Then, from the last Eq.\ of (\ref{baralg}) it follows
that $\bar{D}$ is 
positive for the expanding tori and negative for the contracting ones,
or 
\be
\bar{H} = |\bar{D}|.
\label{hamilt}
\ee

In the general case, we start from the function
\[
 \bar{D} + a\bar{J}^0 +b\bar{J}^1 + c\bar{J}^2.
\]
As it is only the $D$-part which leads to changes in $\tau$, the Hamiltonian
that evolves towards the future corresponding to the above function is
\[
  \bar{H} = \mbox{sign}(\bar{D})(\bar{D} + a\bar{J}^0 +b\bar{J}^1 +
  c\bar{J}^2).
\]
Eqs.\ (\ref{baralg}) lead to
\[
  \bar{H} = \mbox{sign}({\mathbf x}\cdot{\mathbf p})[{\mathbf
  x}\cdot{\mathbf p} + 
  a(x_1p_2 - x_2p_1) + bx_2\sqrt{{\mathbf p}\cdot{\mathbf p}} -
  cx_1\sqrt{{\mathbf p}\cdot{\mathbf p}}].
\]
Let us change the variables $x_1$, $x_2$, $p_1$ and $p_2$ to $\bar{D}$,
$\bar{J}^0$, $p$, $\alpha$; we obtain
\be
  \bar{H} = |\bar{D}|(1 + b\sin\alpha - c\cos\alpha) +
  \mbox{sign}(\bar{D})\bar{J}^0(a + b\cos\alpha + c \sin\alpha).
\label{positivity}
\ee We can see that $\bar{H}$ is unbounded from below except for the
case that $a = b = c = 0$, because $\bar{J}^0$ can take on an
arbitrary values independently of $\bar{D}$ and $\alpha$. Hence,
(\ref{hamilt}) is the only one from the three-dimensional family of
possible candidates for a Hamiltonian that is bounded from below (and
even positive).  This is, of course, nothing but an intriguing
observation: there is no a priori reason for the generator of the time
evolution, even if it evolves towards the future, to be bounded from
below or positive, unless it plays simultaneously another role, for
example that of the total energy of the system. We also observe that
the dynamics simplifies strongly if we choose (\ref{hamilt}) in
comparison with all other candidates: $\bar{J}^\mu$ become time
independent, and $\bar{P}_\mu$ just scale with time. The next comment
is that the choice (\ref{hamilt}) leads to the dynamics that has been
obtained by Moncrief \cite{M2}. Finally, it is easy to see that there
will be no problem to define the quantum mechanical operator $\hat{H}$
from $\bar{H}$, if the operators $\hat{D}$ and $\hat{J}^\mu$ are
given, because $\hat{D}$ will commute with all $\hat{J}$'s. The
corresponding problem of ranges will be automatically solved, if we
define $|\hat{H}|$ by the spectral theorem.

\section{The physical representation}
\label{sec:repre}
The physical representation of the algebra so(2,3) would map
each element of the algebra to a linear operator on a common invariant dense
domain ${\mathcal K}_0$ in a Hilbert space ${\mathcal K}$; the map $R$ must
satisfy the following conditions:
\ben
\item
  $R$ is linear, $R(1) = \mbox{id}$, and
  \[ \frac{\mbox{i}}{\hbar}R(\{X,Y\}) = R(X)R(Y) - R(Y)R(X)\]
  for all $X,Y \in$ so(2,3) on ${\mathcal K}_0$,
\item
  the operators $R(X)$ for all $X \in$ so(2,3) are essentially self-adjoint
  on ${\mathcal K}_0$, 
\item the problem of ranges is satisfactorily solved,
\item
  the operators $R(X)$ for all $X \in$ so(2,3) satisfy algebraic relations
  that go over to 
  (\ref{rel1})--(\ref{rel4}) in the classical limit.
\een

In general, the group method of quantization of an algebra ${\mathbf
g}$ of observables on a symplectic manifold $(M,\Omega)$ is to find a
unitary representation on a Hilbert space ${\mathcal K}$ of the group
$G$ corresponding to the algebra. Then, the generators of the group
action on ${\mathcal K}$ satisfy automatically the conditions 1 and 2,
but a part of the condition 3 ($P_0 < 0$) and the condition 4 can pose
problems.

In this section, we are going to use an old idea of finding the
physical representation by the group way: the Kostant-Kirillov method
of orbits. This method works quite generally for \textit{finite}
systems. Let us briefly describe the steps of the method (for more
detail see \cite{kirill}, \cite{kostant}).

The method of orbits is based on the momentum map $\Pi$ determined by
the algebra of observables ${\mathbf g}$ (in this way, the relations
and ranges are encoded). $\Pi(M)$ is a particular orbit $\omega$ of
$G$ in the linear space ${\mathbf g}^*$ dual to the Lie algebra
${\mathbf g}$, where the group acts via the co-adjoint representation.

The method starts with a choice of a point $F \in \omega$ and with
calculating the stabilizer $G_F \subset G$ of $F$. Then, the
subalgebra ${\mathbf n}_F$ called subordinate to $F$ must be found
satisfying the conditions:
\ben
\item
 $\langle F,[X,Y]\rangle = 0\quad \forall X,Y \in {\mathbf n}_F$,
\item
 codim$_{{\mathbf g}} {\mathbf n}_F = (1/2)$dim $\omega$,
\item
 Pukanszky's condition: let ${\mathbf n}_F^\bot$ be the subspace of
 ${\mathbf g}^*$  that annihilates ${\mathbf n}_F$; then, $F +
 {\mathbf n}_F^\bot 
 \subset \omega$.
\een
One can show that ${\mathbf g}_F \subset {\mathbf n}_F$. The subalgebra
 ${\mathbf n}_F$ generates a subgroup $N_F$ of $G$ and one must find a
 one-dimensional unitary representation $R_{\mathbf n}$ of $N_F$ such that
\[
  R_{\mathbf n}(\exp X) = \exp(\langle F,X\rangle)
\]
in a neighbourhood of the identity of $N_F$. Such a representation
will exist, if Kirillov's symplectic form of $\omega$ is integral (its
integral over any 2-cycle is an integer). The physical representation
is then just the unitary representation of $G$ induced by $N_F$ (see
\cite{kirill}, \cite{B-R}).

In our case, $M = \Gamma_0$ and as the group we take
first SO(2,3). The momentum map $\Pi$ will be described in terms of a
coordinate system in ${\mathbf g}^*$; the coordinate system 
will be associated with the basis that is dual to (\ref{a-basis}); let
the corresponding coordinates be $(\xi_\mu, \zeta_\mu, \theta^\mu,
\delta)$. Then, the momentum map is given by
\be
\left.
\begin{array}{llll}
 \xi_0 = -\sqrt{{\mathbf p}\cdot{\mathbf p}},& \zeta_0 =
-({\mathbf x}\cdot{\mathbf x})\sqrt{{\mathbf p}\cdot{\mathbf p}}, &
\theta^0 = x_1p_2 - x_2p_1, \\  
 \xi_1 = p_1, &
 \zeta_1 = ({\mathbf x}\cdot{\mathbf x})p_1 - 
 2({\mathbf x}\cdot {\mathbf p})x_1, & \theta^1 =
 -x_2\sqrt{{\mathbf p}\cdot{\mathbf p}}, \\
 \xi_2 =
 p_2, & \zeta_2 =
 ({\mathbf x}\cdot{\mathbf x})p_2 - 
 2({\mathbf x}\cdot {\mathbf p})x_2, & \theta^2 
 = x_1\sqrt{{\mathbf p}\cdot{\mathbf p}}, \\
 &&\delta = {\mathbf x}\cdot {\mathbf p}.  
\end{array}
\right\}
\label{mu}
\ee
As the group SO(2,3) has a trivial center, the momentum map is a symplectic
isomorphism and the homogeneous symplectic space
$(\Gamma_0, \Omega_0)$ of $G$ can be identified with
the orbit $\omega$. Then, the Kirillov symplectic form $\Omega_0$ is
exact and so it is trivially integral. 

Let us choose the point $F$ corresponding to the point $u \in
\Gamma_0$ that is given by the values of coordinates $x_1 =
x_2 = 0$, $p_1 =1$ and $p_2 = 0$. From Eqs.\ (\ref{mu}), we calculate the
coordinates of $F$ to be $(-1,1,0,0,0,0,0,0,0,0)$. The map of ${\mathbf g}$
into $T_F\omega$ is given by the values of the vector fields
(\ref{hamvecP})--(\ref{hamvecD}) at the point $u$:
\be
 {\mathcal P}_0 \mapsto -\frac{\partial}{\partial x_1},\quad {\mathcal P}_k
 \mapsto \frac{\partial}{\partial x_k}, 
\label{TFP}
\ee
\be
 {\mathcal Q}_0 \mapsto 0,\quad {\mathcal Q}_k \mapsto 0,
\label{TFQ}
\ee
\be
 {\mathcal J}^0 \mapsto \frac{\partial}{\partial p_2},\quad {\mathcal J}^1
 \mapsto \frac{\partial}{\partial p_2},\quad {\mathcal J}^2 \mapsto
 -\frac{\partial}{\partial p_1},
\label{TFJ}
\ee
\be
 {\mathcal D} \mapsto - \frac{\partial}{\partial p_1},
\label{TFD}
\ee The kernel of the map is, therefore, the subalgebra ${\mathbf
g}_F$ generated by ${\mathcal P}_0 + {\mathcal P}_1$, ${\mathcal J}^0
- {\mathcal J}^1$, ${\mathcal D} - {\mathcal J}^2$ and ${\mathcal
Q}_\mu$, $\mu = 0,1,2$.  The algebra ${\mathbf g}_F$ has to be
extended to ${\mathbf n}_F$. Thus, we have to find a two-dimensional
subspace of ${\mathbf g}/{\mathbf g}_F$ which is invariant with
respect to ${\mathbf g}_F$. ${\mathbf g}/{\mathbf g}_F$ is
four-dimensional; we choose $[{\mathcal P}_1]$, $[{\mathcal P}_2]$,
$[{\mathcal J}^1]$ and $[{\mathcal J}^2]$ as its basis. The action of
$Y \in {\mathbf g}_F$ on ${\mathbf g}/{\mathbf g}_F$ is given by $[X]
\mapsto \pi([X,Y])$, where $X$ is a representant of an element of the
basis of ${\mathbf g}/{\mathbf g}_F$, $[X]$ is the corresponding class
and $\pi$ is the projector from ${\mathbf g}$ to ${\mathbf g}/{\mathbf
g}_F$. A straifgtforward calculation gives
\begin{center}
\begin{tabular}{l||l|l|l|l|l|l} 
&${\mathcal P}_0 + {\mathcal P}_1$ & ${\mathcal J}^0 - {\mathcal J}^1$ &
${\mathcal D} - {\mathcal J}^2$ & ${\mathcal Q}_0$ & ${\mathcal Q}_1$ &
${\mathcal Q}_2$ \\ \hline\hline
$[{\mathcal P}_1]$ & 0 & $[-{\mathcal P}_2]$ & 0 & $[2{\mathcal J}^2]$ &
$[2{\mathcal J}^2]$ & $[-2{\mathcal J}^1]$ \\ \hline
$[{\mathcal P}_2]$ & 0 & 0 & $[-{\mathcal P}_2]$ & $[-2{\mathcal J}^1]$ &
$[2{\mathcal J}^1]$ & $[2{\mathcal J}^2]$ \\ \hline
$[{\mathcal J}^1]$ & $[{\mathcal P}_2]$ & $[{\mathcal J}^2]$ &
$[-{\mathcal J}^1]$ 
& 0 & 0 & 0 \\ \hline
$[{\mathcal J}^2]$ & 0 & 0 & 0 & 0 & 0 & 0 
\end{tabular}
\end{center}
Our task is to find a two-dimensional (in general complex) common
invariant subspace of all six transformations. The abelian subalgebra
generated by ${\mathcal Q}_0$, ${\mathcal Q}_1$ and ${\mathcal Q}_2$
is represented by triangular matrices. They have a common invariant
subspace $T$ spanned by $[{\mathcal J}^1]$ and $[{\mathcal J}^2]$;
there is no complex linear combination $[a{\mathcal P}_1 + b{\mathcal
P}_2]$ that would be mapped by all ${\mathcal Q}_\mu$, $\mu = 0,1,2$,
to a one-dimensional subspace of $T$. Thus, $T$ is the only
two-dimensional invariant subspace of all ${\mathcal Q}_\mu$, $\mu =
0,1,2$. However, this subspace is not invariant with respect to
${\mathcal P}_0 + {\mathcal P}_1$. Hence, no subordinate algebra
${\mathbf n}_F$ exists for the whole group SO(2,3).

In fact, if we just want to have a unitary representation of SO(2,3)
by complex functions on a two-dimensional manifold $M$ (this reflects
the fact that we have two physical degrees of freedom), then such a
representation will determine a definite action of SO(2,3) on $M$ that
will be transitive, or else the representation will not be
irreducible. Then, $M = SO(2,3)/G_M$, where $G_M$ is a stabilizer of a
point of $M$. Thus, SO(2,3) had to admit an 8-dimensional
subgroup. However, there is no such subgroup \cite{B-R}.

There is, however, ${\mathbf n}_F$, if we restrict ourselves to some
subgroup of SO(2,3): the largest are $G_1$, generated by $({\mathcal
P}_0, {\mathcal P}_1, {\mathcal P}_2, {\mathcal J}^0, {\mathcal J}^1,
{\mathcal J}^2, {\mathcal D})$ and $G_2$ generated by $({\mathcal
Q}_0,{\mathcal Q}_1, {\mathcal Q}_2,{\mathcal J}^0,{\mathcal
J}^1,{\mathcal J}^2,{\mathcal D})$.  $\Gamma_0$ is still a space where
$G_1$ or $G_2$ act; they do not act transitively, however: the points
with $x_1 = x_2 = 0$ are invariant with respect to $G_2$ and those
with $y_1 = y_2 = 0$ with respect to $G_1$. These points have to be
cut out in respective cases. Thus, the coordinate patch $(x_k,p_k)$ is
a homogeneous space of $G_1$ and $(y_k,q_k)$ that of $G_2$. The action
of the group $G_1$ on $T^*U$ is the same as that of $G_2$ on $T^*V$ in
the respective coordinates. Let us consider $G_1$ and $T^*{\mathbf
M}^3$ only. In fact, only this case is a group extension of our
original system, namely a $G_1$-extension.

From the corresponding part of the table, we can see immediately that
there are two different invariant subspaces: $T_1$ spanned by
$[{\mathcal P}_1]$ and $[{\mathcal P}_2]$ and $T_2$ spanned by
$[{\mathcal P}_2]$ and $[{\mathcal J}^2]$. It is easy to see that
there are no others. From $T_2$, we obtain the subalgebra ${\mathbf
n}_{2F}$ generated by ${\mathcal P}_0 + {\mathcal P}_1$, ${\mathcal
P}_2$, ${\mathcal J}^0 - {\mathcal J}^1$, ${\mathcal J}^2$ and
${\mathcal D}$; ${\mathbf n}_{2F}$ satisfies the conditions 1 and 2,
but it does not satisfy Pukanszky's condition. Indeed, the subspace
${\mathbf n}_{2F}^\bot$ that annihilates it has the form
$(a,-a,0,b,b,0,0)$ where $(a,b) \in {\mathbf R}^2$. The subset $F +
{\mathbf n}_{2F}^\bot$ is given by $(-1+a,1-a,0,b,b,0,0)$. This will
lie in $\omega$ if the equations
\[
\begin{array}{lll}
 -\sqrt{{\mathbf p}\cdot{\mathbf p}} = -1+a,& p_1 = 1-a,&p_2 = 0, \\
 x_1p_2 - x_2p_1 = b, & -x_2\sqrt{{\mathbf p}\cdot{\mathbf p}} = b,&
 x_1\sqrt{{\mathbf p}\cdot{\mathbf p}} = 0, \\
 {\mathbf x}\cdot{\mathbf p} = 0,&&
\end{array}
\]
have solutions for any $a$ and $b$. However, the first equation
implies that $a < 1$. Thus, this algebra is not admissible. Similar
calculation for $T_1$ shows that the corresponding algebra ${\mathbf
n}_{1F}$ satisfies all three conditions and so it is the only
possibility. Let us concentrate on ${\mathbf n}_{1F}$, which is
generated by ${\mathcal P}_0$, ${\mathcal P}_1$, ${\mathcal P}_2$,
${\mathcal J}^0 - {\mathcal J}^1$ and ${\mathcal D} - {\mathcal J}^2$.

The map of the Lie algebra ${\mathbf g}_1$ of the group $G_1$ into
$T_F\omega$ given by Eqs.\ (\ref{hamvecP}) and
(\ref{hamvecJ0})--(\ref{hamvecD}) sends ${\mathbf n}_{1F}$ on the
subspace $E_F \in T_F\omega$ spanned by the vectors 
\be
\frac{\partial}{\partial x_1},\quad \frac{\partial}{\partial x_2}.
\label{polar}
\ee 
We can use the action ad$^*$ of the group $G_1$ to bring the
subspace from the point $F$ to any other point of $\omega$; this is a
well-defined procedure, because ${\mathbf n}_{1F}$ is invariant with
respect to the stabilizer $G_{1F}$ of $F$. The result is the subspace
spanned by (\ref{polar}) at any point $(x_1,x_2,p_1,p_2)$ of
$\omega$. Indeed, we can use the four one-dimensional subgroups of
$G_1$ generated by ${\mathcal P}_1$, ${\mathcal P}_2$, ${\mathcal
J}^0$ and ${\mathcal J}^2$. The corresponding vector fields given by
Eqs.\ (\ref{hamvecP}), (\ref{hamvecJ0}) and (\ref{hamvecJ2}) describe
the action of these generators on $\omega$; their projections to the
submanifold $x_1 = x_1^0$, $x_2 = x_2^0$ for any constant $x_1^0$ and
$x_2^0$ is independent of $x_1^0$ and $x_2^0$. Thus, the curve defined
by \be x_1 = \gamma_1(t),\quad x_2 = \gamma_2(t),\quad p_1 =
p_1^0,\quad p_2 = p_2^0,
\label{curv1}
\ee 
with the real constants $p_1^0$ and $p_2^0$ will be mapped onto a
curve of the form
\[
  x_1 = \gamma'_1(t),\quad x_2 = \gamma'_2(t),\quad p_1 = p_1^{\prime
  0},\quad
  p_2 = p_2^{\prime 0}  
\]
by any element of the group. Hence, the subspace $E_F$ will be mapped
to the subspace $E_{F'}$ spanned by the vectors (\ref{polar}) at any
point $F'$ of $\omega$ with $x_1 = x_2 = 0$ and $p_1$ and $p_2$
arbitrary. The vector fields (\ref{polar}) (which are now used in
their role of the action of ${\mathcal P}_1$ and ${\mathcal P}_2$) can
easily be integrated; they generate the maps
\[
 (x_1,x_2,p_1,p_2) \mapsto (x_1 + a,x_2 + b,p_1,p_2)
\]
with arbitrary $a$ and $b$. Thus, curves of the form (\ref{curv1}) are
mapped to
\[
 x_1 = \gamma_1(t) + a,\quad x_2 = \gamma_2(t) + b,\quad p_1 = p_1^0,
  \quad p_2 = p_2^0.
\]
Hence, $E_{F'}$ goes over to $E_{F''}$ spanned again by (\ref{polar}) in an
arbitrary point $F''$ of $\omega$. The resulting subbundle of the tangent
bundle is called \textit{polarization}. It is an integrable subbundle, its
integral manifolds $\mathcal{E}$ being given by $p_1 =$ const, $p_2 =$ const. 

At this stage, it is much quicker to guess the form of the operators
representing the Lie algebra that to calculate the representation
according to the general procedure by Kostant-Kirillov.  The unitary
representation of $G_1$ that we are going to construct is induced by
the representation $R_{{\mathbf n}}$ of the subgroup $N_{1F}$ that is
generated by the subalgebra ${\mathbf n}_{1F}$. Thus, the Hilbert space
will be built from complex functions on the homogeneous space
$G_1/N_{1F}$. This may be identified with the manifold
$\bar{\Gamma}_{\mbox{\footnotesize ISO}}/\mathcal{E}$ that is just
${\mathbf R}^2 \setminus \{0\}$ with the coordinates $p_1$ and $p_2$ and
which we have denoted by ${\mathcal P}$ in Sec.\ \ref{sec:extISO}. If we
look at the formula for the induced representation (see e.g.\
\cite{B-R}, P. 479, formula (15)), we can see that there will be three
kinds of terms in the operators representing the Lie algebra of
$G_1$. From the representation of $N_{1F}$, multiplicative terms will
come; they must clearly be multipications by $-\sqrt{{\mathbf p}\cdot{\mathbf
p}}$, $p_1$ and $p_2$ for the operators $\hat{P}_0$, $\hat{P}_1$ and
$\hat{P}_2$. From the action of $G_1$ on the classes $G_1/N_{1F}$,
differential operators come; they must be projections of the vector
fields (\ref{hamvecP}), (\ref{hamvecJ0})--(\ref{hamvecD}) to the space
${\mathcal P}$ multiplied by -i:
\[
 J^0_{\mbox{\footnotesize diff}} = \mbox{i}p_2\frac{\partial}{\partial p_1} 
 -\mbox{i}p_1\frac{\partial}{\partial p_2},
\]
\[
 J^1_{\mbox{\footnotesize diff}} = 
 -\mbox{i}\sqrt{{\mathbf p}\cdot{\mathbf p}}\frac{\partial}{\partial
  p_2},\quad  
 J^2_{\mbox{\footnotesize diff}} =
 \mbox{i}\sqrt{{\mathbf p}\cdot{\mathbf p}}\frac{\partial}{\partial p_1}
\]
\[
 D_{\mbox{\footnotesize diff}} = \mbox{i}p_1\frac{\partial}{\partial p_1} 
 + \mbox{i}p_2\frac{\partial}{\partial p_2}.
\]
Finally, there will be terms coming from the Radon-Nikodym derivative that
will correct the differential operators. Such terms have the general form
\[
 \frac{\mbox{i}}{2\sigma}(\xi_{\mbox{\footnotesize diff}}\sigma),
\]
where $\sigma$ is a quasi-invariant measure on $G_1/N_{1F}$. Different but
equivalent measures will lead to unitarily equivalent representations. A 
choice that strongly simplifies the correction terms is
\[
 \sigma = \frac{1}{\sqrt{{\mathbf p}\cdot{\mathbf p}}}.
\]
Then, finally, the operators must have the form
\be
  \hat{P}_0\psi(p) = -\sqrt{{\mathbf p}\cdot{\mathbf p}}\,\psi(p),\quad
  \hat{P}_k\psi(p) = p_k\psi(p), 
\label{hatP}
\ee
\be
 \hat{J}^0\psi(p) = \mbox{i}p_2\frac{\partial\psi(p)}{\partial p_1}
 - \mbox{i}p_1\frac{\partial\psi(p)}{\partial p_2},
\label{hatJ0}
\ee
\be
 \hat{J}^1\psi(p) = 
 -\mbox{i}\sqrt{{\mathbf p}\cdot{\mathbf p}}\frac{\partial\psi(p)}{\partial
 p_2},\quad  
 \hat{J}^2 = 
  \mbox{i}\sqrt{{\mathbf p}\cdot{\mathbf p}}\frac{\partial\psi(p)}{\partial
 p_1},
\label{hatJ}
\ee
\be
 \hat{D} = \mbox{i}p_1\frac{\partial\psi}{\partial p_1} +
 \mbox{i}p_2\frac{\partial\psi}{\partial p_2} + \frac{\mbox{i}}{2}\psi.
\label{hatD}
\ee 
It is straightforward but tedious to verify that this guessed
operators coincide with those that would follow from the full general
construction of the representation.

An interesting question is, what happend with the relations. For the goup
$G_1$, we have only three relations, and we can take Eqs.\ (\ref{rel1}) and
(\ref{quadrel}). Composing the corresponding operators on a common invariant
domain (say, $C_0^\infty({\mathcal P})$), we obtain easily: 
\bea
 - \hat{P}_0^2 + \hat{P}_1^2 + \hat{P}_2^2 & = & 0,
\label{Psqr} \\ 
 \hat{J}^0\hat{P}_0 + \hat{J}^1\hat{P}_1 + \hat{J}^2\hat{P}_2 & = & 0,
\label{PJ} \\ 
 -\hat{D}^2 - (\hat{J}^0)^2 + (\hat{J}^1)^2 + (\hat{J}^2)^2 & = &
\frac{1}{4}.
\label{Dsqr}
\eea Thus only the last relation has been deformed (of course, there
will be $\hbar^2/4$ at the right hand side, if $\hbar$, which has been
set equal 1, is restored). The left hand sides of Eqs.\
(\ref{Psqr})--(\ref{Dsqr}) are generalized Casimirs of the group $G_1$
in the following sense. Let $E({\mathbf g}_1)$ be the envelopping
algebra of ${\mathbf g}_1$, let ${\mathcal H} = -{\mathcal P}_0^2 +
{\mathcal P}_1^2 + {\mathcal P}_2^2$ represent the constraint and let
${\mathcal I}({\mathcal H})$ be the ideal in $E({\mathbf g}_1)$
generated by ${\mathcal H}$. Then, the classes of the left hand sides
of Eqs.\ (\ref{Psqr})--(\ref{Dsqr}) in $E({\mathbf g}_1)/{\mathcal
I}({\mathcal H})$ commute with all elements of $E({\mathbf
g}_1)/{\mathcal I}({\mathcal H})$.

It is interesting to observe that each state $\psi$ of the Hilbert
space ${\mathcal K}$ of the physical representation must satisfy the
equation
\be
  (- \hat{P}_0^2 + \hat{P}_1^2 + \hat{P}_2^2)\psi = 0,
\label{WDW}
\ee
which follows from Eq.\ (\ref{Psqr}). This is a point where the group
method touches the operator constraint method of quantization, because
Eq.\ (\ref{WDW}) has the form of the operator constraint equation of
our system (for more detail, see \cite{pohl}).

\subsection*{Acknowledgements}
Helpful discussions with J.~Bi\v{c}\'{a}k, K.~Kucha\v{r}, R.~Loll,
V.~Moncrief and L.~Ziewer are gratefully acknowledged. The author is
thankful to National Science Foundation grant PHY9507719, to the
Tomalla Foundation, Zurich and to the Swiss Nationalfonds for
support. Thanks go to the Max-Planck-Institut f\"{u}r
Gravitationsphysik, Potsdam for hospitality and support.


\begin{thebibliography}{99}
\bibitem{carlip1} S.~Carlip, ``Six ways to quantize
  (2+1)-dimensional gravity'' in \textit{General Relativity and
  Relativistic Astrophysics}, Proceedings, edited by R.~B.~Mann and
  R.~G.~McLenaghan. World Scientific, Singapore, 1994. Pp. 215--234.
\bibitem{loll1} R.~Loll, J. Math.\ Phys.\ \textbf{36} (1995) 6494.
\bibitem{dirac1} P.~A.~M.~Dirac, \textit{Lectures on Quantum Mechanics}.
  Yeshiva University Press, New York, 1964.
\bibitem{cordoba} K.~V.~Kucha\v{r}, ``Canonical Quantum Gravity'' in
  \textit{General Relativity and Gravitation 1992}; edited by
  R.~J.~Gleiser, C.~N.~Kozameh and O.~M.~Moreschi. IOP Publishing, 
  Bristol, 1993. Pp. 119--150.
\bibitem{PH5} P.~H\'{a}j\'{\i}\v{c}ek, Class.\ Quant.\ Grav.\ \textbf{13}
 (1996) 1353--1375.
\bibitem{bergmann1} P.~G.~Bergmann, Rev.\ Mod.\ Phys.\ \textbf{33} (1961)
  510--514.
\bibitem{bluebook} A.~Ashtekar, \textit{Lectures on Non-Perturbative
  Canonical Gravity}. World Scientific, Singapore, 1991.
\bibitem{kuchS} K.~V.~Kucha\v{r}, J. Math.\ Phys.\ \textbf{22} (1981) 2640
\bibitem{AT} Anderson and C.~Torre, Commun.\ Math.\ Phys.\
  \textbf{176} (1996) 479.
\bibitem{ishamG} C.~J.~Isham, ``Topological and Global Aspects of
  Quantum Theory,'' in \textit{Relativity, Groups and Topology II}.
  Ed.\ by B.~C.~DeWitt and R.~Stora. Elsevier, New York, 1984.
\bibitem{R-group} C.~Rovelli, Nuovo Cimento B \textbf{100} (1987) 343.
\bibitem{dirac2} P.~A.~M.~Dirac, Rev.\ Mod.\ Phys.\ \textbf{21} (1949)
  392.
\bibitem{PH1} P.~H\'{a}j\'{\i}\v{c}ek, J.\ Math.\ Phys.\ \textbf{36} (1995)
  4612--4638.
\bibitem{carlip} E.~Witten, Nucl.\ Phys.\ \textbf{B311} (1988) 46.
\bibitem{mart} S.~P.~Martin, Nucl.\ Phys.\ \textbf{B327} (1989) 178.
\bibitem{loll2} R.~Loll, Class.\ Quantum Grav.\ \textbf{12} (1995) 1655. 
\bibitem{pohl} K.~Pohlmayer, Commun.\ Math.\ Phys.\ \textbf{114} (1988) 351. 
\bibitem{kostant} B.~Kostant in \textit{Lectures in Modern Analysis and
 Applications. III} Edited by C.~J.~Taam. Berlin, Springer, 1970.
\bibitem{kirill} A.~A.~Kirillov \textit{Elements of the Theory of
  Representations}. Berlin, Springer, 1976.
\bibitem{anderss} L.~Andersson, private communication.
\bibitem{LM} J.~Louko and D.~Marolf, Class.\ Quantum Grav.\
  \textbf{11} (1994) 311.
\bibitem{M1} V.~Moncrief, J. Math.\ Phys.\ \textbf{30} (1989) 2907.
\bibitem{M2} V.~Moncrief, J. Math.\ Phys.\ \textbf{31} (1990) 2978.
\bibitem{M-M} N.~Manojlovic and A.~Mikovic, Nucl.\ Phys.\
\textbf{B385} (1992) 571.
\bibitem{terras} A.~Terras \textit{Harmonic Analysis on Symmetric Spaces and
  Applications. I,II} Berlin, Springer, 1985/88.
\bibitem{HE} S.~W.~Hawking and G.~F.~R.~Ellis, \textit{The Large Scale
  Structure of Space-Time}. Cambridge University Press, Cambridge, 1973.
\bibitem{P-R} R.~Penrose and W.~Rindler, \textit{Spinors and
  Space-Time}. Cambridge University Press, Cambridge, 1986. Vol.\ II,
  P. 297.
\bibitem{B-R} A.~O.~Barut and R.~R\c{a}czka, \textit{Theory of Group
 Representations and Applications}. Warsaw, Polish Scientific
 Publishers, 1980.


\end{thebibliography}
\end{document}